\documentclass[10pt,journal,compsoc]{IEEEtran}
\ifCLASSOPTIONcompsoc
  \usepackage[nocompress]{cite}
\else
  \usepackage{cite}
\fi
\ifCLASSINFOpdf
\else
\fi
\usepackage{cite}
\usepackage{amsmath,amssymb,amsfonts}
\usepackage{algorithm}
\usepackage{algorithmic}
\usepackage{amsthm}
\usepackage{color}
\usepackage{graphicx}
\usepackage{subfigure}
\usepackage{csquotes}
\usepackage{textcomp}
\usepackage{url}
\def\BibTeX{{\rm B\kern-.05em{\sc i\kern-.025em b}\kern-.08em
		T\kern-.1667em\lower.7ex\hbox{E}\kern-.125emX}}
\newtheorem{lemma}{\hskip\parindent\bf{Lemma}}[section]
\newtheorem{definition}{\hskip\parindent\bf{Definition}}[section]
\newtheorem{theorem}{\hskip\parindent\bf{Theorem}}[section]
\newtheorem{corollary}{\hskip\parindent\bf{Corollary}}[section]
\newtheorem{example}{\hskip\parindent\bf{Example}}[section]
\newsavebox{\sembox}
\newlength{\semwidth}
\newlength{\boxwidth}

\newcommand{\ms}{M_s}

\newcommand{\type}{\gamma}
\newcommand{\seq}{\varepsilon}

\newcommand{\specI}{\mathcal{I}}
\newcommand{\specX}{\mathcal{X}}
\newcommand{\specY}{\mathcal{Y}}

\newcommand{\ivs}{\textsf{\upshape{ivs}}}

\newcommand{\para}{\textsf{\upshape{par}}}

\newcommand{\pred}{\textsf{\upshape{pre}}}

\newcommand{\preall}{\textsf{\upshape{ans}}}

\newcommand{\sucd}{\textsf{\upshape{suc}}}

\newcommand{\sucall}{\textsf{\upshape{des}}}

\newcommand{\resultR}{\mathcal{R}}
\newcommand{\resultF}{\mathcal{F}}
\newcommand{\clause}{\mathcal{C}}

\newcommand{\p}[1]{\textsf{\upshape{par}}(#1)}

\newcommand{\colornew}[1]{\textsf{\upshape{ivs}}(#1,s)}

\begin{document}
%
\title{Response Time Bounds for Typed DAG Parallel Tasks on Heterogeneous Multi-cores}
%
%
%
%

\author{Meiling~Han,~\IEEEmembership{}
        Nan~Guan,~\IEEEmembership{}
        Jinghao~Sun,~\IEEEmembership{}
        Qingqiang~He,~\IEEEmembership{}
        Qingxu~Deng,~\IEEEmembership{}
        and~Weichen~Liu,~\IEEEmembership{}
        \IEEEcompsocitemizethanks{	\IEEEcompsocthanksitem This paper is under submission to a journal.	
        	\IEEEcompsocthanksitem M. Han,J. Sun and Q. Deng are with Northeastern University, Shenyang 110819, China. Part of the work by M. Han was performed during her visit at  The Hong Kong Polytechnic University \protect\\
        	E-mail: neu$\_$hml@163.com
        	\IEEEcompsocthanksitem N. Guan and Q. He are with The Hong Kong Polytechnic University, Hong Kong. Weichen Liu is with Nanyang Technological University, Singapore.}
        \thanks{Manuscript received-; revised-.}}
\markboth{Journal of \LaTeX\ Class Files,~Vol.~14, No.~8, August~2015}%
{Shell \MakeLowercase{\textit{et al.}}: Bare Demo of IEEEtran.cls for Computer Society Journals}
%



\IEEEtitleabstractindextext{%
\begin{abstract}
Heterogenerous multi-cores utilize the strength of different architectures for executing particular types of workload, and usually offer higher performance and energy efficiency. In this paper, we study the worst-case response time (WCRT) analysis of \emph{typed} scheduling of parallel DAG tasks on heterogeneous multi-cores, where the workload of each vertex in the DAG is only allowed to execute on a particular type of cores. The only known WCRT bound for this problem is grossly pessimistic and suffers the \emph{non-self-sustainability} problem. In this paper, we propose two new WCRT bounds. The first new bound has the same time complexity as the existing bound, but is more precise and solves its \emph{non-self-sustainability} problem. The second new bound explores 
more detailed task graph structure information to greatly improve the precision, but is computationally more expensive. We prove that the problem of 
computing the second bound is strongly NP-hard if the number of types in the system is a variable, and develop an efficient algorithm which has polynomial time complexity if the number of types is a constant. 
Experiments with 
randomly generated workload show that our proposed new methods are significantly more precise than the existing bound while having good scalability.
\end{abstract}

}

\maketitle

\IEEEdisplaynontitleabstractindextext

%
\IEEEpeerreviewmaketitle


\IEEEraisesectionheading{\section{Introduction}\label{sec:intru}}

Multi-cores are more and more widely used in real-time
systems, to meet rapidly increasing requirements in
performance and energy efficiency. To fully utilize the computation capacity of multi-cores, software should be properly parallelized. 
A representation that can model a wide range of parallel software is the DAG (directed acyclic graph) task model, where each vertex represents a piece of sequential workload and each edge represents the precedence relation between two vertices.
Real-time scheduling and analysis of DAG parallel task models {have} raised many new challenges over traditional real-time scheduling theory with sequential tasks, and {have} become an increasingly hot research topic in recent years.

Many modern multi-cores adopt heterogeneous architectures. 
{Examples include Zynq$-$7000 \cite{Zynq7000} and OMAP1/OMAP2 \cite{OMAP} that integrate CPU and DSP on the same chip, and 
	the Tegra processors \cite{Tegra} that integrate CPU and GPU on the same chip.}
Heterogenerous multi-cores utilize specialized processing capabilities to handle particular computational tasks, which
usually offer higher performance and energy efficiency.
For example, \cite{Venkat2014} showed that a heterogeneous-ISA chip multiprocessor can outperform the best same-ISA homogeneous architecture by as much as $21\%$ with $23\%$ energy savings and a reduction of $32\%$ in energy delay product.

In this paper, we consider real-time scheduling of \emph{typed DAG tasks} on heterogeneous multi-cores, where each vertex is explicitly bound to execute on a particular type of cores. Binding 
code segments of the program to a
certain type of cores is common practice in software development on heterogeneous multi-cores and is supported by mainstream parallel programming frameworks and operating systems. For example, in OpenMP \cite{openmp} one can use the \textbf{proc$\_$bind} clause to specify 
the mapping of threads to certain processing cores. 
{In OpenCL\cite{OpenCL}, one can use the \textbf{clCreateCommandQueue} function to create a command queue to certain devices. In CUDA \cite{cuda}, one can use the \textbf{cudaSetDevice} function to set the following executions to the target device.} 
%
%
%

The target of this paper is to bound
the \emph{worst-case response time} (WCRT) for typed DAG tasks.

To the best of our knowledge, the only known WCRT bound for the considered problem model  was presented in an early work \cite{Jaffe1980} (called \textbf{OLD-B}), 
which is not only grossly pessimistic, but also suffers the \emph{non-self-sustainability} problem\footnote{
	By a non-self-sustainable
	analysis method, 
	a system decided to be schedulable 
	may be decided to be unschedulable when the system parameters become \enquote{better}. We will discuss this issue in more details in Section \ref{s:first}. }. 
In this paper we develop two new response time bounds to address these problems:

\begin{itemize}
	\item \textbf{NEW-B-1}, which  dominates \textbf{OLD-B} in analysis precision with the same time  complexity and solves its non-self-sustainability problem. 
	
	\item \textbf{NEW-B-2}, which significantly improves
	the analysis precision by exploring more detailed task graph stucture information. \textbf{NEW-B-2} is more precise, but also more difficult to compute.  
	\begin{itemize}
		\item We prove the problem of computing \textbf{NEW-B-2} to be strongly NP-hard if the number of types 
		is a \emph{variable}.
		\item We develop an efficient algorithm to compute \textbf{NEW-B-2} with polynomial time complexity if the number of types
		is a \emph{constant}.
	\end{itemize}

\end{itemize}

Experiments with randomly generated parallel tasks 
show that the new WCRT bounds proposed in this paper can greatly improve the analysis precision.
This paper focuses on analysis of a single typed DAG task, but our results are also meaningful to general system setting with multiple recurrent typed DAG tasks.
On one hand, the results of this paper are directly applicable to multiple tasks under scheduling algorithms where a subset of cores are assigned to each individual parallel task (e.g., federated scheduling {\cite{Li2014,Baruah2015,Baruah2015a,Beckert2017,Li2017a}}).
On the other hand, 
the analysis of \emph{intra-task} interference addressed in this paper is 
a necessary step towards the analysis
for scheduling algorithms where different tasks interfere with each other (e.g., global scheduling {\cite{Bonifaci2013,Baruah2014,Melani2015}}).

%
\section{Preliminary }

\subsection{\textbf{Task Model}}\label{s:model}
We consider a \emph{typed DAG task} $G = (V, E, \type, c)$ to be executed on a heterogeneous multi-core platform with different types of cores.
$S$ is the set of core types (or \emph{types} for short), and for each $s \in S$ there are $M_s$ cores of this type ($M_s \geq 1$). 
$V$ and $E$ are the set of vertices and edges in $G$.
Each vertex $v \in V$ represens 
a piece of code segment to be sequentially executed.
Each edge $(u,v) \in E$ represents the precedence relation between vertices $u$ and $v$. 
The \emph{type function} $\type: V \times S$ defines the \emph{type} of each vertex,  i.e.,
$\type(v) = s$, where $s \in S$, represents vertex $v$ must be executed on cores of type $s$. 
The \emph{weight function} $c: V \times \mathbb{R}_0^+$ defines the worst-case execution time (WCET) of each vertex, i.e., $v$ executes for at most $c(v)$ time units (on cores of type $\type(v)$). 

\begin{figure}
	\centering
	\includegraphics[scale=0.85]{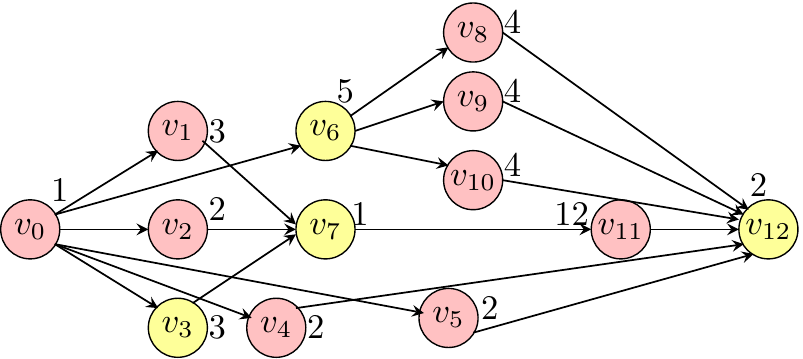}
	\caption{A typed DAG task $\tau$ with two types. }
	\label{fig:dag}
\end{figure}

If there is an edge $(u, v) \in E$, $u$ is a \emph{predecessor} of $v$, and $v$ is a \emph{successor} of $u$. 
If there is a path in $G$ from $u$ to $v$, $u$ is an \emph{ancestor} of $v$ and $v$ is a \emph{descendant} of $u$. We use $\pred(u)$, 
$\sucd(u)$, $\preall (u)$ and $\sucall (u)$ to denote the set of predecessors, 
successors, ancestors and descendants of $u$, respectively.
Without loss of generality, we assume $G$ has a unique source vertex $v_{src}$ (which has no predecessor) and a unique sink vertex $v_{snk}$ (which has no successor)\footnote{In case $G$ has multiple source/sink vertices, one can add a dummy source/sink vertex to make it compliant with our model.}.
We use $\pi \in G$ to denote $\pi$ is a path in $G$.
A path $\pi = \{\tau_1, \cdots, \tau_k  \}$ is a \emph{complete path}
iff its first vertex $\tau_1$ 
is the source vertex of $G$
and last vertex $\tau_k$ is the sink vertex.
We use $vol(G)$ to denote the total WCET of $G$ and $vol_s(G)$ the total WCET of vertices of type $s$: 
\[
vol(G) = \sum_{u\in V} c(u),~~ vol_s(G) = \sum_{u\in V \wedge \type(u) = s} c(u).
\]
The length of a path $\pi$ is denoted by $len(\pi)$ and $len(G)$ represents the length of the longest path in $G$:
\[
len(\pi) = \sum_{u \in \pi} c(u),~~len(G) = \max_{\pi \in G} \{ len(\pi) \}.
\]
\begin{example}\label{ex:runseqence}
	Figure \ref{fig:dag} illustrates a typed DAG task with 
	two types of vertices (type $1$ marked by yellow and 
	type $2$ marked by red). The WCET of vertex is annotated by the number next to the vertex.
	And we can compute that $vol(G)=45$, $vol_1(G)=11$ and $vol_2(G)=34$. For a path $\pi=\{v_0,v_1,v_7,v_{11},v_{12}\}$, the length is $len(\pi)=19$.	
\end{example}	

\begin{figure} 
	\centering 
	\subfigure[An execution sequence where each vertex executes for its WCET.]{\label{fig:runseq1} 
		\includegraphics[scale=0.33]{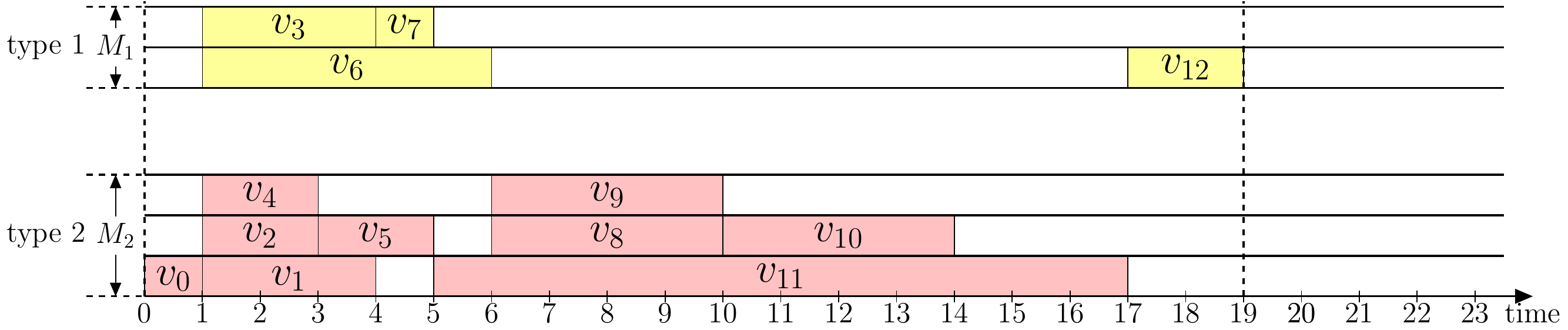}}  
	\subfigure[An execution sequence where some vertices executes shorter than their WCET.]{ \label{fig:runseq2} 
		\includegraphics[scale=0.33]{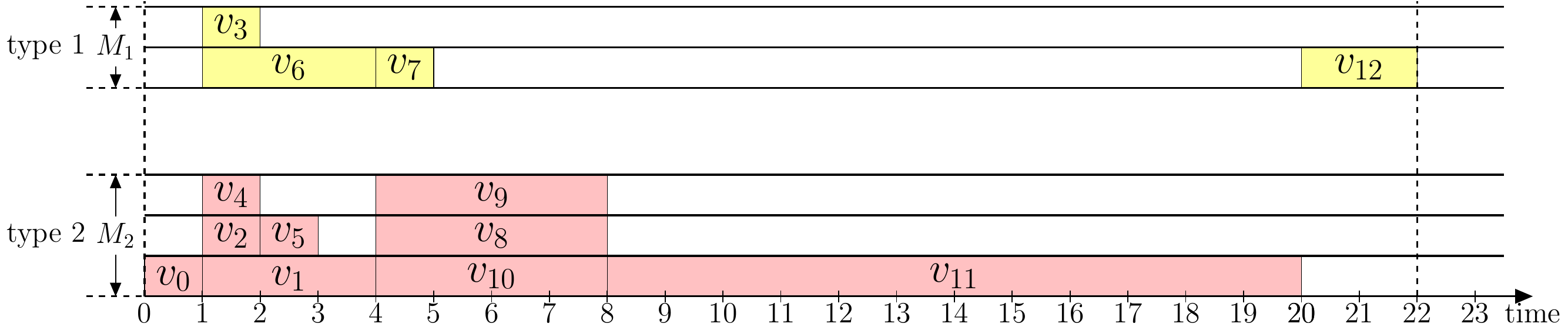}}
	\caption{Two possible execution sequences the task in Figure\ref{fig:dag} executed on a platform with $M_1 =2$ and $M_2 = 3$.} 
	\label{fig:runseq} 
\end{figure}

\subsection{\textbf{Runtime Behavior}}\label{s:runtime}

A vertex is \emph{eligible} for execution when all of its predecessors have finished. Without loss of generality, we assume the source vertex of $G$ is eligible for execution at time $0$. 
The typed DAG task $G$ is scheduled on the heterogeneous multi-core platform by a \emph{work-conserving} scheduling algorithm:

\begin{definition}
	Under a \emph{work-conserving} scheduling algorithm, an eligible vertex of type $s$ must be executed if there
	are available cores of type $s$.
\end{definition}

%
We do not put any other constraints to scheduling algorithms except the {work-conserving} constraint. There are many possible instances 
of work-conserving scheduling algorithms, e.g., the \emph{list scheduling} {\cite{Graham1966}} algorithm. The results of this paper are applicable to any work-conserving scheduling algorithm.


\textbf{Execution Sequence.} At runtime,
the vertices of $G$ execute at certain time according to the scheduling algorithm. We call a trace describing 
which vertex executes at which time points an  \emph{execution sequence} of $G$.
Given a  scheduling algorithm, $G$ may generate different execution sequences. This is because, (1) the scheduling algorithm may have nondeterminism (the scheduler may behave differently in the same situation) and (2) each vertex may execute for shorter  than its WCET.
For example, Figure \ref{fig:runseq1} shows an execution 
sequence where each vertex executes for its WCET, while 
Figure \ref{fig:runseq2} shows another execution sequence where
some vertices execute for shorter than their WCET but lead to a larger response time.
In an execution sequence $\seq$, we use $f_{\seq}(v)$ to denote the finish time of vertex $v$. For simplicity, we omit the subscript and only use  $f(v)$ to denote $v$'s finish time when the execution sequence is 
clear from the context.
%


%


\textbf{Response Time.}
The \emph{response time} of $G$ in an execution sequence is the finish time of the sink vertex, and the WCRT of $G$, denoted by $R(G)$, is the maximum among the response times of all possible execution sequences. 
The target of this paper is to derive safe upper bounds for the WCRT of $G$. Note that the WCRT of $G$ is not necessarily achieved by the execution sequence in which each vertex executes for its WCET (even if there is only one type in the system) \cite{Graham1966,Graham1969}.
Therefore, one can not obtain the WCRT of $G$ by simply simulating the execution of $G$
using the WCET, but has to (explicitly or implicitly) analyze all the possible execution sequences of $G$.

\subsection{\textbf{Existing WCRT Bound}}\label{s:classicmethod}

To our best knowledge, the only known
WCRT upper bound for the considered model 
was developed in an early work \cite{Jaffe1980}:
\begin{theorem}[\textbf{OLD-B}]
	\label{th:bwcrt}
	The WCRT of  $G$ is bounded by:
	\begin{equation}	\label{eq:bwcrt}
	R(G) \leq \left (1 - \frac{1}{
		\displaystyle \max_{s \in S} \{M_s \}} \right ) \times len(G) + \sum_{s \in S} \frac{vol_s(G)}{\ms}.
	\end{equation}
\end{theorem}
This bound can be computed in $O(|V| + |E|)$ time \cite{dasgupta2006algorithms}.
Although \textbf{OLD-B} was originally derived for the \emph{list scheduling} algorithm \cite{Graham1966}, it applies to all work-conserving scheduling algorithms. 
When there is only one type, it degrades to the classical response time bound for \emph{untyped} DAG  tasks {\cite{Graham1969}}:
\begin{equation*}	
R(G) \leq len(G)  + \frac{vol(G) - len(G)}{M}.
\end{equation*}

%
%

%
%
%
%
%
%


%
%
%

\section{The First New WCRT Bound}\label{s:first}

\textbf{OLD-B} is not only pessimistic but
also suffers the problem of being   \emph{non-self-sustainable} with respect to processing capacity.
More specifically, the value of the WCRT bound in
(\ref{eq:bwcrt}) may increase when the
number of cores (of some type) increases, as witnessed by the following example.
%
\begin{example}
	For the task $G$ in Figure \ref{fig:dag}, we can calculate its $vol_1(G)=11$, $vol_2(G)=34$ and $len(G)=19$. 
	Suppose $M_1=2$ and $M_2=3$, we obtain a WCRT bound by \textbf{OLD-B} as $29.5$. However, if we increase $M_1$ to $20$, the bound is increased to $29.9\dot{3}$. 
\end{example}

Note that the actual WCRT of $G$ will
\emph{not} increase when more cores are used. The phenomenon shown above is merely the problem of the bound \textbf{OLD-B} itself rather than the system behavior. As pointed out in \cite{Baker2009}, 
the \emph{self-sustainability} property is important in incremental and interactive design process, which is typically used in the design of real-time systems and in the evolutionary development of fielded systems.

In this section we will develop a new WCRT bound, which is not only more precise than \textbf{OLD-B} (with the same time complexity), but also self-sustainable.
We start with introducing some useful concepts.
%

\begin{definition}
	\label{de:sg}
	The \emph{scaled graph} $\hat{G} =  (V, E, \hat{c}, \gamma)$ of $G = (V, E, c, \gamma)$ has the same topology ($V$ and $E$) and type function $\gamma$ as $G$, but a different weight function $\hat{c}$:
	\[
	\forall v\in V: \hat{c}(v) = c(v) \times (1 - 1/M_{\type(v)}).
	\]
\end{definition}


%
%

\begin{definition}
	A \emph{critical path} $\pi = \{\tau_1, \cdots, \tau_k \}$ of an execution sequence of $G$
	is a \emph{complete} path of $G$ satisfying the following condition:
	\[
	\forall \tau_i \in \pi \setminus \{ \tau_1\}: f(\tau_{i-1})
	= \max_{u \in \pred (\tau_i)} \{ f(u) \},
	\]
	where $f(v)$ is the finish time of $v$ in this execution sequence.
\end{definition}

For example, a complete path $\pi = \{v_0,v_1,v_7,v_{11},v_{12}\}$ is the critical path for the execution sequence shown in Figure \ref{fig:runseq1}, while a complete path $\pi' = \{v_0,v_2,v_7,v_{11},v_{12}\}$
is \emph{not} a critical path of this execution sequence 
since the $v_2$'s finish time is not the latest among all
the predecessors of $v_7$.

%

A task $G$ may generate (infinitely) many different 
execution sequences at runtime, and it is in general unknown which complete path in $G$ is the critical path that leads to the WCRT. In the following, we assume an \emph{arbitrary} complete path $\pi = \{\tau_1, \cdots, \tau_k \}$ to be a critical path, and derive upper bounds for the response time of this particular critical path. Then by getting the maximum bound among all possible paths in $G$, we can safely bound the WCRT of $G$.

We divide $[0, f(\tau_k))$ into $k$ segments
$[0, f(\tau_1))$, $[f(\tau_{1}), f(\tau_{2}))$, $\cdots$, $[f(\tau_{k-1}), f(\tau_{k}))$. 
For each $ 1 < i \leq k$, we define
\[
I_i =  f(\tau_i) - f(\tau_{i-1}),
\]
and let $I_1 =  f(\tau_1)$. We define
\begin{itemize}
	\item $x_i$: the accumulative length of time intervals in $I_i$ during which $\tau_i$ is executing;
	\item $y_i$: the accumulative length of time intervals in $I_i$ during which $\tau_i$ is
	\emph{not} executing.
\end{itemize}
%
%
%
%
\begin{figure*}
	\centering
	\includegraphics[width=6.4in,height=2.2in]{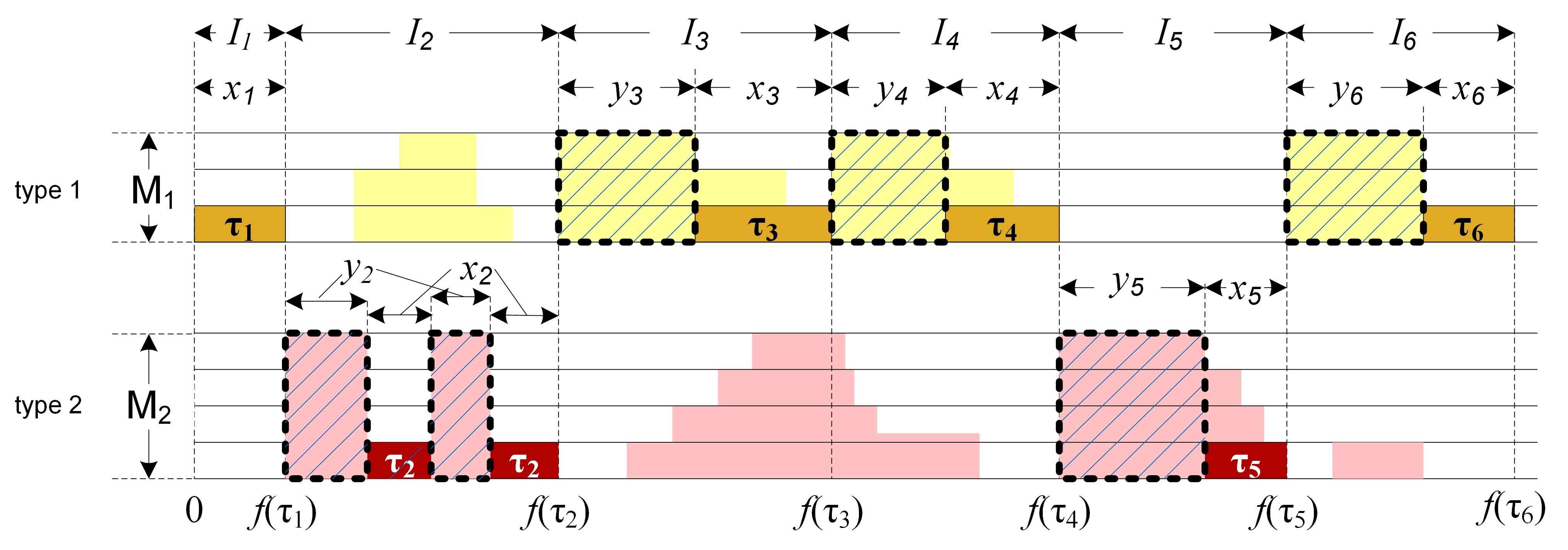}
	\caption{Illustration of $x_i$, $y_i$ and $I_i$.}
	\label{fig:window}
\end{figure*}
Obviously, $I_i = x_i + y_i$. Figure \ref{fig:window} illustrates $x_i$, $y_i$ and  $I_i$. In general the time intervals counted in $x_i$ or $y_i$ may not be continuous (e.g., $[f(\tau_1), f(\tau_2))$ in Figure \ref{fig:window}).
We further define 
\[
\specI_s = \sum_{ \substack{\tau_i \in \pi \wedge \\ \type(\tau_i) = s }} I_i, ~~	\specX_s = \sum_{\substack{\tau_i \in \pi \wedge \\ \type(\tau_i) = s } } x_i, ~~
\specY_s = \sum_{\substack{\tau_i \in \pi \wedge \\ \type(\tau_i) = s } } y_i,
\]
and we know $\specI_s = 	\specX_s + 	\specY_s$.	

%
%

\begin{lemma}\label{lm:xy}
	Let $\pi = \{\tau_1, \cdots, \tau_k \}$ be a critical path of an arbitrary execution sequence of $G$, then $\specX_s$ and 	$\specY_s$ can be bounded by
	\vspace{-2mm}
	\begin{align}
	\specX_s & \leq \sum_{\substack{\tau_j \in \pi \wedge \\ \type(\tau_j) = s }} c(\tau_{j}), \label{e:specxbound} \\
	\specY_s & \leq  \left (
	vol_s(G) - 	 \sum_{\substack{\tau_j \in \pi \wedge \\ \type(\tau_j) = s }} c(\tau_{j}) \right ) / M_s. \label{e:specybound}
	\end{align}
	
\end{lemma}

\begin{proof}
	The proof of 
	(\ref{e:specxbound}) is trivial. In the following, we focus on the proof of (\ref{e:specybound}). 
	By the definition of critical path, we know all the predecessors of $\tau_{i}$ have finished by time $f(\tau_{i-1})$. Therefore, when $\tau_i$ is not executing in $[f(\tau_{i-1}), f(\tau_{i}))$, all the cores of 
	type $\type(\tau_{i})$ must be occupied by vertices 
	of type $\type(\tau_{i})$ \emph{not} on the critical path.
	Since the total workload of vertices of type $s$ that are \emph{not} on the critical path is at most 
	\[
	vol_s(G) - 	 \sum_{\substack{\tau_j \in \pi \wedge \\ \type(\tau_j) = s }} c(\tau_{j})
	\]
	and the number of cores of type $s$ is $M_s$, 
	the accumulated length of time intervals 
	during which the vertices of type $s$ on $\pi$
	are not executing is bounded by (\ref{e:specybound}).
\end{proof}

\begin{theorem}[\textbf{NEW-B-1}]
	\label{th:newwcrt1}
	The WCRT of $G$ is bounded by:
	\vspace{-2mm}
	\begin{equation}\label{e:newwcrt1}
	R(G) \leq len(\hat{G}) + \sum_{s \in S} \frac{vol_s(G)}{\ms},
	\vspace{-2mm}
	\end{equation}
	where $\hat{G}$ is the scaled graph of $G$.	
\end{theorem}

\begin{proof}
	By (\ref{e:specxbound}), (\ref{e:specybound}) and $\specI_s = \specX_s + \specY_s$ we have
	\begin{align*}
	\specI_s & \leq \frac{vol_s(G)}{M_s}  + \sum_{\substack{\tau_i \in \pi \wedge \\ \type(\tau_i) = s}}  \left(1 - 1/M_s \right)  c(\tau_{i})   \\
	\sum_{s \in S} \specI_s &	\leq	\sum_{s \in S} 
	\frac{vol_s(G)}{M_s} + 
	\sum_{s \in S} \sum_{\substack{\tau_i \in \pi \wedge \\ \type(\tau_i) = s}}  \left(1 - 1/M_s \right)  c(\tau_{i}) \\
	\sum_{s \in S} \specI_s &	\leq	\sum_{s \in S} 
	\frac{vol_s(G)}{M_s} + 
	\sum_{\substack{\tau_i \in \pi}}   \hat{c}(\tau_{i}) ~ \emph{\textrm{//by the definition of~}} \hat{c}(\tau_{i})
	\end{align*}
	$\sum_{s \in S} \specI_s$ is the response time
	of the execution sequence with critical path $\pi$.
	Since $G$ and $\hat{G}$ have the same topology, 
	$\pi$ is also a complete path in $\hat{G}$, so 
	$ \sum_{\substack{\tau_i \in \pi}}   \hat{c}(\tau_{i}) $ is bounded by $len(\hat{G})$. Therefore, $R(G)$ is bounded by (\ref{e:newwcrt1}).
\end{proof}

%
%
%
%
%
%
%
%

We can compute $\sum_{s \in S} vol_s(G) / M_s$
and construct $\hat{G}$ based on $G$
in $O(|V|)$ time, and compute $len(\hat{G})$ in $O(|V| + |E|)$ time \cite{dasgupta2006algorithms}. Therefore, the overall time complexity to compute \textbf{NEW-B-1} is $O(|V| + |E|)$, which is the same as \textbf{OLD-B-1}. 
By comparing the two bounds
we can conclude:
\begin{corollary}
	\textbf{\textrm{NEW-B-1}} \emph{strictly dominates} \textbf{\textrm{OLD-B-1}}. 
\end{corollary}

Finally, we can easily see the bound in 
(\ref{e:newwcrt1}) is decreasing with respect to each $M_s$, so we can conclude:
\begin{corollary}
	\textbf{\textrm{NEW-B-1}} is \emph{self-sustainable} with respect to each $M_s$.
\end{corollary}

\section{The Second New WCRT Bound}

Our first new WCRT bound \textbf{NEW-B-1} is more precise than \textbf{OLD-B}, but still very pessimistic.
The source of its pessimism comes from the step of bounding $\specY_s$.
Intuitively, the bound of $\specY_s$ in (\ref{e:specybound}) 
is derived assuming that the 
workload of vertices not on the critical path are all executed in the shaded areas in Figure \ref{fig:window}.
However, in reality much workload of $G$ may actually be executed outside these shaded areas.
Therefore, the length of $\specY_s$ is significantly over-estimated in (\ref{e:specybound}) .

In this section, we introduce the second 
new WCRT bound \textbf{NEW-B-2}, which
eliminates workload of vertices that cannot be executed in the shaded area, and thus reduce the pessimism in bounding $\specY_s$.


\subsection{\textbf{WCRT Bound}}

%
%

\begin{definition}
	\label{de:para}
	For each vertex $v\in V$, $\p v$ denotes the set of vertices that have the same type as $v$ but are neither ancestors nor descendants of $v$:
	\begin{equation*}
	\p v=\{u | {u \in  V} \wedge {\type (u)=\type (v)} \wedge {u \notin (\preall (v)\cup \sucall (v))} \}.
	\end{equation*}
\end{definition}



\begin{definition}
	\label{de:pcor}
	Let $\pi = \{\tau_1,\cdots,\tau_{k}\}$ be a critical path, $\ivs(\pi, s)$
	is defined as
	\begin{equation}\label{eq:color}
	\ivs(\pi, s) =
	\bigcup_{\substack{\tau_{i}\in \pi \wedge \\ \type ({\tau_{i}})=s}} \p {\tau_{i}}.
	\end{equation}
\end{definition}

\begin{example}
	Assume $\pi = \{v_0,v_1,v_7,v_{11},v_{12}\}$ is a critical path of the task in Figure \ref{fig:dag}. We have $\p {v_0}=\emptyset$, $\p {v_1}=\{v_2,v_4,v_5,v_8,v_9,v_{10}\}$, $\p {v_7}=\{v_6\}$, $\p {v_{11}}=\{v_4,v_5,v_8,v_9,v_{10}\}$, $\p {v_{12}}=\emptyset$, $\ivs (\pi,1)=\{v_6\}$  and $\ivs (\pi,2)=\{v_2,v_4,v_5,v_8,v_9,v_{10}\}$.
\end{example}

Intuitively, $\ivs(\pi, s)$
is the set of vertices of type $s$ that are not on the critical path but can actually interfere with vertices of type $s$ on the critical path (i.e., can be executed in the shaded area in Figure \ref{fig:window}). Therefore, $\specY_s$ can be bounded more precisely as stated in the following Lemma.

\begin{lemma}\label{l:xy}
	Let $\pi = \{\tau_1, \cdots, \tau_k\}$ be a critical path of an arbitrary execution sequence of $G$, then $\specY_s$ is bounded by
	\begin{equation}
	\specY_s  \leq  \frac{\ivs(\pi, s)}{M_s} \label{e:newspecybound}
	\end{equation}
	
\end{lemma}

\begin{proof}
	
	To prove the lemma, it is sufficient to prove that
	at any time instant in $[f(\tau_{i-1}), f(\tau_{i}))$ when $\tau_{i}$
	is \emph{not} executing, all the cores of type $\type (\tau_{i})$	must be executing vertices in $\ivs(\pi, \type (\tau_i))$.	
	We prove this by contradiction. Assuming that at a time instant $t \in [f(\tau_{i-1}), f(\tau_{i}))$ when $\tau_{i}$
	is not executing, there exists a core of type $\type (\tau_i)$
	which is \emph{not} executing vertices in $\ivs(\pi, \type (\tau_i))$, 
	then one of the following two cases must be true:
	\begin{itemize}
		\item This core is idle at $t$.
		Since $\pi$ is a critical path, we know all the predecessors of $\tau_{i}$ have finished by time $f(\tau_{i-1})$, so $\tau_i$ is eligible for execution at $t$,
		and thus this core cannot be idle at $t$. Therefore, this case is impossible.
		
		%
		
		\item This core is executing a vertex $u \notin \ivs(\pi, \type (\tau_i))$ at $t$. First we know $\type (u) = \type (\tau_i)$, and since
		$u \notin \ivs(\pi, \type (\tau_i))$, by the definition of $\ivs$ 
		we know $u$ must be a predecessor or a successor of $\tau_i$, so we discuss two cases:
		
		\begin{itemize}
			\item $u$ is a predecessor of $\tau_i$. 
			Since $\pi$ is a critical path, we know all the  predecessors of $\tau_{i}$ have finished by time $f(\tau_{i-1})$, so a predecessor of $\tau_i$ cannot start execution after $f(\tau_{i-1})$, which contradicts that $u$ is executing at a time instant $t$ after $f(\tau_{i-1})$.

			\item $u$ is a successor of  $\tau_i$. A successor of 
			$\tau_i$ cannot start execution before $f(\tau_{i})$, so this is also a contradiction.
		\end{itemize} 
		Therefore, this case is also impossible.
	\end{itemize}
	In summary, both cases are impossible, so the assumption must be false and the lemma is proved.
\end{proof}


%


%

\begin{theorem}[\textbf{NEW-B-2}]
	\label{th:wcrt}
	The WCRT of $G$ is bounded by
	\begin{equation}
	\label{eq:wcrt}
	R(G) \leq \max_{\pi\in G} \{ \widetilde{R}(\pi) \}
	\end{equation}	
	where	
	\begin{equation}\label{e:eachR}	
	\widetilde{R}(\pi) = len(\pi)+\sum_{s\in S} \sum_{v\in \ivs(\pi, s)}c(v)/\ms
	\end{equation}
\end{theorem}

\begin{proof}
	
	By the same idea as the proof of Theorem 	\ref{th:newwcrt1} but using the new bound
	(\ref{e:newspecybound}) for $\specY_s$ instead of (\ref{e:specybound}), we can get
	\[
	\sum_{s \in S} \specI_s \leq	len(\pi)+\sum_{s\in S} \sum_{v\in \ivs(\pi, s)}c(v)/\ms	
	\]
	
	$\sum_{s \in S} \specI_s $ is the response time of the execution sequence with $\pi$ being the critical path. 
	Finally, by getting the maximum bound 
	for all complete paths (assumed to be the critical path), the theorem is proved.
\end{proof}	

Comparing with \textbf{NEW-B-1}, \textbf{NEW-B-2} uses a more precise upper bound of $\specY_s$, so we have
\begin{corollary}
	\textbf{NEW-B-2} \emph{strictly dominates} \textbf{NEW-B-1}. 
\end{corollary}
The bound in (\ref{eq:wcrt})
is decreasing with respect to each $M_s$, so
\begin{corollary}
	\textbf{{NEW-B-2}} is \emph{self-sustainable} with respect to each $M_s$.
\end{corollary}
%
%
%
%
%

\subsection{\textbf{Strong NP-Hardness}}\label{ss:nph}

\begin{figure}
	\centering
	\includegraphics[scale=0.63]{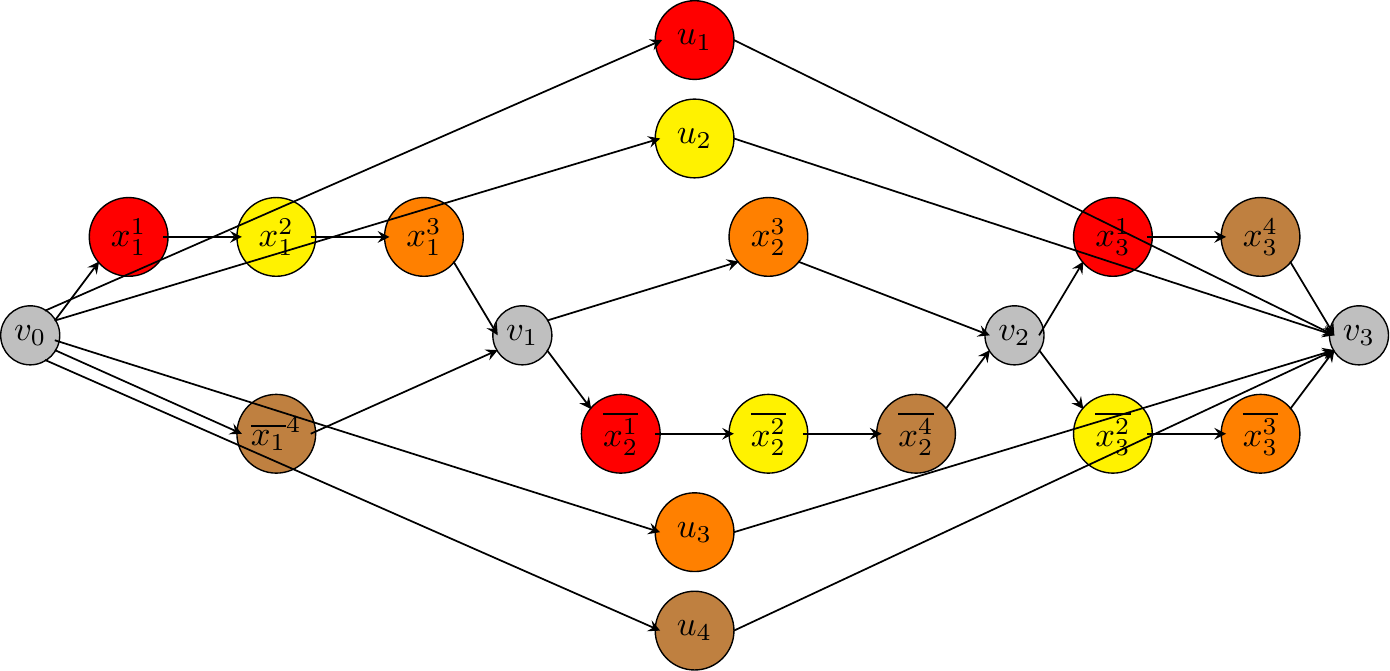}
	\caption{
		The constructed typed DAG task $G$ 
		for 3-SAT problem instance $\clause = \clause_1 \wedge \clause_2 \wedge \clause_3 \wedge \clause_4$, where
		$\clause_1 =x_1 \vee \overline{x_2} \vee x_3$, $\clause_2=x_1\vee \overline{x_2} \vee \overline{x_3}$, $\clause_3=x_1\vee x_2 \vee \overline{x_3}$, $\clause_4=\overline{x_1}\vee \overline{x_2} \vee x_3$. }
	\label{fig:cdag}
\end{figure}
\textbf{{NEW-B-2}} requires to compute the maximum of $\widetilde{R}(\pi)$ among all paths in the graph $G$. It is computationally intractable to explicitly enumerate all the paths, the number of which is exponential.
Can we develop efficient algorithms of (pseudo-)polynomial complexity to compute $\max_{\pi\in G} \{ \widetilde{R}(\pi) \}$? 

Unfortunately, this is impossible unless \textbf{P} = \textbf{NP}.
\begin{theorem}
	\label{th:nph}
	The problem of 
	computing  $\max_{\pi\in G} \{ \widetilde{R}(\pi) \}$ is strongly NP-hard.
\end{theorem}

\begin{proof}
	We will prove the theorem by showing that even a simpler problem of verifying whether $ \max_{\pi\in G} \{ \widetilde{R}(\pi) \}$ is larger than a given value $\omega$ is strongly NP-hard, which is proved by a reduction from the 3-SAT problem.

	Let $\clause$ be an arbitrary instance of the 3-SAT problem, which has
	$m$ clauses $ \clause_1\bigwedge \clause_2 \bigwedge \cdots \bigwedge \clause_m$
	and $n$ variables $\{x_1,x_2,\cdots,x_n\}$.
	Each clause $\clause_r$, $1\leq r\leq m$, consists of three literals, and each literal is a variable or the negation of a variable. 
	We construct a typed DAG $G$ corresponding to the 3-SAT instance as follows:
	\begin{itemize}
		\item We first construct $n+1$  vertices 
		$\{v_0, \cdots, v_n\}$ of type $s_0$ with $c(v_0) = \cdots = c(v_n) = 1$. $v_0$ is 
		the source vertex of $G$ and $v_n$ is the sink vertex.
		
		\item For each clause $\clause_r$, we construct a vertex $u_r$ of type $s_r$ with $c(u_r) = 1$,
		as well as two edges $(v_0, u_r)$ and $(u_r, v_n)$. 
		\item For each variable $x_i$, we construct two paths from $v_{i-1}$ to $v_{i}$:
		\begin{itemize}
			\item \textbf{Positive path},
			which includes a vertex 
			$x_i^{r}$ of type $s_r$ if and only if clause $\clause_r$ includes a literal $x_i$.
			
			\item \textbf{Negative path}, which includes a vertex 		$\overline{x_i^{r}}$ if and only if clause $\clause_r$ includes a literal $\overline{x_i}$.
		\end{itemize}
		The WCET of each vertex on these two paths is $\frac{1}{mn + 1}$.
	\end{itemize}
	
	Note that there are in total $m+1$ types in the above constructed DAG. Finally, we set $M_0 = M_1 = \cdots = M_m = 1$ and $\omega = m + n+1$. The above construction is  polynomial  as there are no more than $m + n + 1 + 2 nm$ vertices in the constructed graph.
	For illustration, an example of the above construction is given in Figure \ref{fig:cdag}.

	In the following we prove that the 3-SAT problem instance $\clause$
	is satisfiable if and only if the bound  $\max_{\pi\in G} \{ \widetilde{R}(\pi) \}$ of the above constructed graph is strictly greater than $ m + n + 1$.
	
	%
	
	First, a complete path that leads to the largest $\widetilde{R}(\pi)$ must be one of those traversing $v_0, \cdots, v_n$.
	The choice between the positive and negative path between $v_{i-1}$ and $v_{i}$ corresponds to the choice between assigning $1$ or $0$ to variable $x_i$ in the 3-SAT problem.
	
	Since each vertex $u_r$ is neither an ancestor nor a descendant of any vertex on paths traversing $v_0, \cdots, v_n$, $u_{r}$ is included in $\ivs(\pi, \type(u_r))$ if and only if the path $\pi$ contains at least one vertex of type $\type(u_r)$.
	This corresponds to that $\clause_r$ is satisfied only if it 
	contains at least one literal assigned with value $1$.
	Therefore, we can conclude that all vertices $u_r$ are 
	included in the corresponding $\ivs(\pi, \type(u_r))$  
	if and only if all clauses contain at least one literal assigned with value $1$, i.e., the 3-SAT problem instance 
	$\clause$ is satisfiable.
	
	Therefore, the second item of RHS of (\ref{e:eachR}) equals $m$ if and only if $\clause$ is satisfiable.
	Moreover, 
	there are at most $mn$ vertices corresponding to the positive and negative values of the variables along any path traversing $v_0, \cdots, v_n$, so their total WCET must be in the range $(0, \frac{mn}{mn+1})$. Therefore,
	the 
	length $len(\pi)$ of any such path $\pi$ must be in the range $(n+1, n+1+\frac{mn}{mn+1})$, and thus in the range $(n+1, n+2)$.
	Therefore, $\max_{\pi\in G}$ is larger than $m+n+1$ if and only if all vertices $u_r$ are included in the corresponding $\ivs(\pi, \type(u_r))$, i.e., the 3-SAT problem instance $\clause$ is satisfiable.
\end{proof}


\subsection{\textbf{Computation Algorithm}}\label{s:newmethod}

The construction in the above strong NP-hardness proof uses $m+1$  different types (where $m$ is the number of clauses in  3-SAT).
In realistic heterogeneous multi-core platforms, the number of core types is usually not very large. 
Will the problem of computing  
$\max_{\pi\in G} \{ \widetilde{R}(\pi)\}$ remain NP-hard if the number of types is a bounded constant? 
In the following, we will present an algorithm to compute $\max_{\pi\in G} \{ \widetilde{R}(\pi)\}$ with complexity ${O(|V|^{|S|+2})}$, which shows that the problem is actually in \textbf{P} if the number of types is a constant.

We first describe the intuition of our algorithm. Instead of explicitly enumerating all the possible paths, our algorithm will
use abstractions to represent paths in the graph searching procedure. 
More specifically, a path starting from the source vertex of $G$ and ending at some vertex $\tau_i$ is abstractly represented by a {tuple} $\langle \tau_i, \Delta(\tau_i), \resultR(\tau_i) \rangle$, 
where $\Delta(\tau_i)$ and $\resultR(\tau_i)$ are defined in Definition 
\ref{def:Delta} and \ref{def:resultR} in the following.
The tuple will be updated when the path is extended from $\tau_i$ to its successor $\tau_{i+1}$, and eventually 
when the path is extended to the sink vertex, 
$\resultR(\tau_i)$ is the $\widetilde{R}(\pi)$ for this path.
The algorithm starts with a single tuple corresponding to the path consisting only the source vertex and repeatedly extends the paths until they all reach the sink vertex, then the maximal  $\widetilde{R}(\pi)$ 
among all the kept tuples is the desired bound
$\max_{\pi\in G} \{ \widetilde{R}(\pi) \}$. 
The abstraction is compact, so that many different path histories ending with the same vertex can be represented by a single abstraction and the total number of abstractions generated in the computation is polynomially bounded.

\begin{definition}
	\label{def:Delta}
	For a path $\pi = \{ \tau_1, \cdots, \tau_k \}$ in $G$ and a type $s \in S$, we define 
	\begin{equation*}
	\delta(\tau_i, s)  =  \begin{cases}
	\tau_i, &  s = \type(\tau_i) \\
	\bot, &  s \neq \type(\tau_1) \wedge  i = 1 \\
	\delta(\tau_{i \textrm{-}1}, s), &  s \neq \type(\tau_i)  \wedge 2 \leq i \leq k 
	%
	%
	\end{cases}
	\end{equation*}
	and 
	\begin{equation}\label{e:delta}
	\Delta(\tau_i) = \{ \delta(\tau_i, s)  ~|~ s \in S \}
	\end{equation}
\end{definition}

Intuitively, $\delta(\tau_i, s)$ is the vertex on path $\pi$ that is the closest to $\tau_i$ among all the vertices of type $s$.

\begin{example}
	In the typed task in Figure \ref{fig:dag}, there are $3$ paths from $v_0$ to $v_7$. For the path $\pi_1=\{v_0,v_1,v_7\}$, one can derive $\delta(v_7,1)=\{v_7\}$ and $\delta(v_7,2)=\{v_1\}$. For the path $\pi_2=\{v_0,v_3,v_7\}$, one can derive $\delta(v_7,1)=\{v_7\}$ and $\delta(v_7,2)=\{v_0\}$.
\end{example}

\begin{definition}\label{def:resultR}
	For a path $\pi = \{ \tau_1, \cdots, \tau_k \}$ in $G$ we define:
	\begin{equation}\label{e:resultR}
	\resultR(\tau_i) =  
	\begin{cases}
	c(\tau_1), & i=1 \\
	\displaystyle \resultR(\tau_{i\textrm{-}1})+c(\tau_i)+ \!\!
	\sum_{v\in \varphi(\tau_{i})} \frac{c(v)}{M_{\type(\tau_{i})}}, 
	&  2 \leq i \leq k
	\end{cases}
	\end{equation}
	where
	\begin{equation*}
	\varphi(\tau_{i}) = \p {\tau_{i}}  \setminus \p{ \delta(\tau_{i\textrm{-}1})}
	\end{equation*}
	(since $\delta(\tau_{i\textrm{-}1})$ may be $\bot$, we let $\para(\bot) = \emptyset$ for completeness).
\end{definition}

%
%
%
%
%

\begin{lemma}\label{l:resultR}
	For a complete path $\pi = \{\tau_1, \cdots, \tau_k\}$, it always holds
	\begin{equation}\label{eq:resultR}
	\resultR(\tau_k) = 
	\widetilde{R}(\pi). 
	\end{equation}
\end{lemma}

%
%

\begin{proof}
	The proof goes in two steps: (1) Rewrite $\resultR(\tau_i)$ into a non-recursive form $\resultR'(\tau_i)$ and prove $\forall \tau_i \in \pi: 		\resultR'(\tau_i) = \resultR(\tau_i)$, and (2)  Prove $\resultR'(\tau_k) = 
	\widetilde{R}(\pi)$.

	We first define $\resultR'(\tau_i)$ as follows:
	\begin{equation*}\label{eq:resultRp}
	\resultR'(\tau_i)
	= \left \{ \begin{array}{ll}
	c(\tau_1) & i =1\\
	\displaystyle	\sum_{j = 1}^i c(\tau_j) + \sum_{s \in S} 
	\sum_{\substack{v \in \varphi(\tau_j) \wedge \\
			\type(\tau_j) = s \wedge j \leq i}}
	\frac{c(v)}{M_s}   & 2 \leq i \leq k		 
	\end{array}
	\right. 
	\end{equation*}	
	
	Now we prove $\forall \tau_i \! \in \! \pi \!: 		\resultR'(\tau_i) = \resultR(\tau_i)$
	by induction:
	
	\begin{itemize}
		\item \textbf{Base case}. Both $\resultR(\tau_1)$ and $\resultR'(\tau_1)$ equal $c(\tau_1)$, so the claim holds for the base case with $i=1$.
		\item \textbf{Inductive step}. Suppose $\resultR(\tau_{i\textrm{-}1}) = \resultR'(\tau_{i\textrm{-}1})$,
		we want to prove $\resultR(\tau_{i}) = \resultR'(\tau_{i})$. First, we know
		\begin{equation}\label{e:abstract-1}
		\!\!\!\!\!\!\!\!\!\!\!\!\!\! \sum_{v \in \varphi(\tau_i)} \frac{c(v)}{M_{\type(\tau_i)}}
		+ \sum_{s \in S} 
		\!\!\!\!\sum_{\substack{v \in \varphi(\tau_j) \wedge \\
				\type(\tau_j) = s \wedge j \leq i-1}} \!\!\!\!\!\!\!\!
		\frac{c(v)}{M_s} =  \sum_{s \in S} 
		\!\!
		\sum_{\substack{v \in \varphi(\tau_j) \wedge \\
				\type(\tau_j) = s \wedge j \leq i}}\!\!\!\!\!
		\frac{c(v)}{M_s}
		\end{equation}
		For simplicity we let 
		\[
		\resultF(\tau_i) = \sum_{s \in S} 
		\!\!
		\sum_{\substack{v \in \varphi(\tau_j) \wedge \\
				\type(\tau_j) = s \wedge j \leq i}}\!\!\!\!\!
		\frac{c(v)}{M_s}
		\]
		so (\ref{e:abstract-1}) can be rewritten as 
		\begin{align*}
		& \sum_{v \in \varphi(\tau_i)} \frac{c(v)}{M_{\type(\tau_i)}} + \resultF(\tau_{i\textrm{-}1}) = \resultF(\tau_{i}) \\
		\Leftrightarrow \!\!&
		\sum_{v \in \varphi(\tau_i)} \!\!\! \frac{c(v)}{M_{\type(\tau_i)}} \!+\! \resultF(\tau_{i\textrm{-}1}) \!+\!\! \sum_{j=1}^{i} c(\tau_j) = 
		\resultF(\tau_{i}) \!+\!\! \sum_{j=1}^{i} c(\tau_j)  \\
		\Leftrightarrow \!\!&
		\sum_{v \in \varphi(\tau_i)} \frac{c(v)}{M_{\type(\tau_i)}} + \resultR'(\tau_{i\textrm{-}1}) + c(\tau_i) =  \resultR'(\tau_{i}) \\
		\Leftrightarrow \!\!&
		\sum_{v \in \varphi(\tau_i)} \frac{c(v)}{M_{\type(\tau_i)}} + \resultR(\tau_{i\textrm{-}1}) + c(\tau_i) =  \resultR'(\tau_{i}) \\
		\Leftrightarrow \!\! & ~~~
		\resultR(\tau_{i}) =  \resultR'(\tau_{i}) 
		\end{align*}
		
	\end{itemize}
	By now, we have proved $\forall \tau_i \in \pi: 	\resultR'(\tau_i) = \resultR(\tau_i)$. In the following we prove $\resultR'(\tau_k) = 
	\widetilde{R}(\pi)$.

	Let $\pi^s=\{\tau^s_{1},\cdots,\tau^s_{h}\}$ be the subsequence of $\pi$ containing all vertices in $\pi$ with type $s$. By the definition of $\varphi({\tau_j})$:
	\begin{equation}\label{eq:resultR-2}
	\sum_{\substack{v \in \varphi({\tau_j}) \\ \wedge \type({\tau_j} ) = 
			s \\ \wedge j \leq k  }}  \!\!\!\! c(v) 
	= \!\!\!\! \!\! \sum_{v\in \p {\tau^s_{1}}} \!\!\!\! c(v) + 
	\!\!\!\!
	\sum_{\substack{v\in \p {\tau^s_{2}} \\ \setminus \p {\tau^s_{1}}}} \!\!\!\! c(v)+\cdots + \!\!\!\!\sum_{\substack{v\in \p {\tau^s_{h}} \\ \setminus \p {\tau^s_{h \textrm{-}1}}}} \!\!\!\!c(v)
	\end{equation}
	
	In the following we will prove 
	\begin{equation}\label{e:ivsequalsub}
	\sum_{v\in \colornew \pi} \!\!\!\! c(v) = \!\!\!\! \!\! \sum_{v\in \p {\tau^s_{1}}} \!\!\!\! c(v) + 
	\!\!\!\!
	\sum_{\substack{v\in \p {\tau^s_{2}} \\ \setminus \p {\tau^s_{1}}}} \!\!\!\! c(v)+\cdots + \!\!\!\!\sum_{\substack{v\in \p {\tau^s_{h}} \\ \setminus \p {\tau^s_{h \textrm{-}1}}}} \!\!\!\!c(v)
	\end{equation}
	If (\ref{e:ivsequalsub}) is true, then by (\ref{e:ivsequalsub})
	and (\ref{eq:resultR-2}) we have
	\[
	\sum_{\substack{v \in \varphi({\tau_j}) \\ \wedge \type({\tau_j} ) = 
			s \\ \wedge j \leq k  }}  c(v)  = \sum_{v\in \colornew \pi}c(v) 
	\]
	by which $\resultR'(\tau_k) = 
	\widetilde{R}(\pi)$ is proved. 
	
	In the following, we focus on proving  (\ref{e:ivsequalsub}). We use LHS and RHS to represent the left-hand side
	and right-hand side of (\ref{e:ivsequalsub}), respectively. In the following, we will prove that both LHS $\leq$ RHS and LHS $\geq$ RHS
	hold.
	\begin{enumerate}
		\item \textbf{LHS $\leq$ RHS}. 
		This is proved by combining the following two claims:
		\begin{enumerate}
			\item  \textbf{Any $c(v)$ counted in LHS is also counted in RHS.}
			If $c(v)$ is counted in LHS, then $v$ must be in some $\para(\tau_i^s)$, and by the definition of $\ivs(\pi, s)$,
			we know $v$ must be also in $\colornew \pi$, so we can conclude that all the $c(v)$ counted in LHS {are} also counted in RHS.

			\item \textbf{Each $c(v)$ is counted in  LHS at most once.} Suppose this is not true, then there exists some $v$ such that $v\in \p {\tau^s_{i}} \setminus \p {\tau^s_{i\textrm{-}1}}$ and $v\in \p {\tau^s_{j}} \setminus \p {\tau^s_{j\textrm{-}1}}$
			for some $j < i-1$, so 
			it must be the case that
			\[	v \in  \p {\tau^s_{i}}  ~\wedge~ 		v \notin  \p {\tau^s_{i \textrm{-}1}} ~\wedge~ 		v \in  \p {\tau^s_{j}} \]

			By $v \in  \p {\tau^s_{i}}$, we know 
			$v$ is not a descendant of $\tau^s_{i}$, and since 
			$\tau^s_{i \textrm{-}1}$ is a predecessor of $\tau^s_{i}$,
			$v$ is not a descendant of $\tau^s_{i \textrm{-}1}$ either.
			On the other hand, by $v \in  \p {\tau^s_{j}}$
			we know $v$ is not an ancestor of $\tau^s_{j}$, and since $\tau^s_{i \textrm{-}1}$ is a descendant of $\tau^s_{j}$,  $v$ is not an ancestor of $\tau^s_{i-1}$. In summary, $v$ is neither a descendant nor an ancestor of $\tau^s_{i \textrm{-}1}$, and $v$ has the same type as $\tau^s_{i \textrm{-}1}$, which contradicts $v \notin  \p {\tau^s_{i \textrm{-}1}}$.

		\end{enumerate}

		\item \textbf{RHS $\leq$ LHS}. 
		It is obvious that each $c(v)$ is counted at most once in the RHS, so it suffices to prove that 
		each $c(v)$ counted in the RHS is also counted in LHS.
		Since $v \in \colornew \pi$, $v$ must be in some $\p{\tau_i^s}$. {Suppose $i$ is the smallest index such that $v \in \p{\tau_i^s}$ but $v \notin \p{\tau_{i-1}^s}$},
		then $c(v)$ is counted in item $\sum_{v\in \p {\tau^s_{i}} - \p {\tau^s_{i\textrm{-}1}}}  c(v)$ (if $i=1$ then $c(v)$ is counted in $\sum_{v\in \p {\tau^s_{1}} }  c(v)$). 	
	\end{enumerate}
	In summary, we have proved both RHS $\geq$ LHS and RHS $\leq$ LHS, so (\ref{e:ivsequalsub}) is true.
\end{proof}

By Lemma \ref{l:resultR}, we know that by using the 
abstract tuple  $\langle \tau_i, \Delta(\tau_i), \resultR(\tau_i) \rangle$ to extend the path, eventually, we can precisely compute 		$\widetilde{R}(\pi)$. Therefore, we can use 
$\langle \tau_i, \Delta(\tau_i), \resultR(\tau_i) \rangle$ as the abstraction of paths to perform graph searching. All the paths corresponding to the same tuple
can be abstractly represented by a single tuple instead of recording each of them individually.
Actually, even paths corresponding to different tuples can be merged together during the graph searching procedure by the \emph{domination} relation among tuples defined as follows:

\begin{definition}\label{d:domination}
	Given two tuples $\langle v, \Delta_1(v), \resultR_1(v) \rangle$ and $\langle v, \Delta_2(v), \resultR_2(v) \rangle$ with the same vertex $v$, 
	$\langle v, \Delta_1(v), \resultR_1(v) \rangle$ \emph{dominates} $\langle v, \Delta_2(v), \resultR_2(v) \rangle$,
	denoted by 
	$$\langle v, \Delta_1(v), \resultR_1(v) \rangle
	\succcurlyeq
	\langle v, \Delta_2(v), \resultR_2(v) \rangle,
	$$ if both of the following conditions are satisfied:
	\begin{enumerate}
		\item $\resultR_1(v) \geq \resultR_2(v)$
		\item $\forall s$: either $\delta_1(v, s) = \bot$
		or 			
		\vspace{-1.5mm}
		\begin{align*}
		& (\delta_1(v, s) \!\neq\! \bot) ~\wedge~ 			 (\delta_2(v, s) \!\neq\! \bot) ~\wedge~ \\
		& ~~~~~~~~~~~~~~~~~~~~~(\para(\delta_1(v, s)) \cap \sucall( \delta_2(v,s)) \!=\! \emptyset)
		\end{align*}
	\end{enumerate}
\end{definition}

{Suppose $v$ is a successor of $u$. Given a 
	tuple $\langle u, \Delta(u), \resultR(u) \rangle$ for vertex $u$ and a tuple  $\langle v, \Delta(v), \resultR(v) \rangle$ for vertex $v$}, we use
\[
\langle u, \Delta(u), \resultR(u) \rangle \mapsto \langle v, \Delta(v), \resultR(v) \rangle
\]
to denote that $\langle v, \Delta(v), \resultR(v) \rangle$ is generated based on $\langle u, \Delta(u), \resultR(u) \rangle$ according to  (\ref{e:delta}) and (\ref{e:resultR}).

\begin{lemma}\label{l:domination}
	Suppose $v$ is a successor of $u$, and
	\begin{align*}
	\langle u, \Delta_1(u), \resultR_1(u) \rangle & \mapsto \langle v, \Delta_1(v), \resultR_1(v) \rangle \\
	\langle u, \Delta_2(u), \resultR_2(u) \rangle & \mapsto \langle v, \Delta_2(v), \resultR_2(v) \rangle
	\end{align*}
	if $\langle u, \Delta_1(u), \resultR_1(u) \rangle \succcurlyeq \langle u, \Delta_2(u), \resultR_2(u) \rangle$, then we must have
	$
	\langle v, \Delta_1(v), \resultR_1(v) \rangle \succcurlyeq \langle v, \Delta_2(v), \resultR_2(v) \rangle$.
\end{lemma}

\begin{proof}
	We start with proving $\resultR_1(v) \geq \resultR_2(v)$.
	Assume $w_1=\delta_1(u, \type(v))$ and $w_2=\delta_2(u, \type(v))$, then \[
	\varphi_1(v) = \para(v) \setminus \para(w_1)
	\]
	\[
	\varphi_2(v) = \para(v) \setminus \para(w_2)
	\]
	In the following we prove $ \varphi_1(v)  \supseteq  \varphi_2(v) $. 
	By condition 2) in Definition \ref{d:domination}, one of the following two cases must hold
	\begin{enumerate}
		\item $w_1 = \bot$ 
		\item $(w_1 \!\neq\! \bot) ~\wedge~ 			( w_2 \!\neq\! \bot) ~\wedge~ (\para(w_1) \cap \sucall(w_2) \!=\! \emptyset)$
	\end{enumerate}
	If $w_1= \bot$, then
	$\para(w_1) = \emptyset$, so $ \varphi_1(v)  \supseteq  \varphi_2(v) $ is obviously true.
	In the following we focus on the second case $$(w_1 \!\neq\! \bot) ~\wedge~ 			 (w_2 \!\neq\! \bot) ~\wedge~ (\para(w_1) \cap \sucall(w_2) \!=\! \emptyset).$$
	Suppose $ \varphi_1(v)  \nsupseteq  \varphi_2(v) $ , then there must exist an $x$ s.t. 
	$$x \in \varphi_2(v)  \wedge x \notin \varphi_1(v),$$ by which we know any following conditions must hold  
	\begin{align}
	x & \in \p {v} \label{e:31}\\
	x &\in \p {w_1} \label{e:32}\\
	x &\notin \p {w_2} \label{e:33}
	\end{align}
	By (\ref{e:31}) we have $x \notin \preall(v) \cup \sucall (v)$
	and thus
	\begin{equation}\label{e:34}
	x \notin \preall(v)
	\end{equation}
	By (\ref{e:33}), we have
	\begin{equation}\label{e:35}
	x \in \preall(w_2) \cup \sucall (w_2).
	\end{equation}
	Since $w_2=\delta_2(u, \type(v))$ and $v \in \sucd(u)$, we know $w_2\in \preall (v)$, and thus
	\begin{equation}\label{e:36}
	\preall (w_2) \subset \preall (v)
	\end{equation}	     
	By (\ref{e:34}),  (\ref{e:35}) and  (\ref{e:36}) we know
	$x \in  \sucall (w_2)$, which contradicts with $\p {w_1}\cap \sucall(w_2)=\emptyset$, so the assumption $\varphi_1(v)  \nsupseteq  \varphi_2(v)$ must be false, i.e., $ \varphi_1(v)  \supseteq  \varphi_2(v) $ must be true.  Therefore 
	\[ \sum_{w \in \varphi_1(v)} \frac{c(w)}{M_{\type(v)}} \geq \sum_{w \in \varphi_2(v)} \frac{c(w)}{M_{\type(v)}},\]
	and since $\resultR_1(u) \geq \resultR_2(u) $ , we have 
	\[
	\resultR_1(u) + c(v) +
	\sum_{w \in \varphi_1(v)} \frac{c(w)}{M_{\type(v)}} \geq 
	\resultR_1(u) + c(v) 
	\sum_{w \in \varphi_2(v)} \frac{c(w)}{M_{\type(v)}}\]	  	  by which we have $\resultR_1(v) \geq \resultR_2(v)$. 
	
	Next we prove the second condition of the domination relation is also true. We discuss two cases of $s$:
	\begin{itemize}
		\item $s = \type(v)$. In this case  
		$\delta_1(v, s) = \delta_2(v, s) = v$, so we have $
		(\delta_1(v, s) \neq \bot ) \wedge 
		(\delta_2(v, s) \neq \bot )	
		\wedge
		(\para(\delta_1(v, s)) \cap 
		\sucall(\delta_2(v, s)) = \emptyset)
		$
		
		\item $s \neq \type(v)$. In this case,
		$\delta_1(v, s) = \delta_1(u, s)$
		and 	 	$\delta_2(v, s) = \delta_2(u, s)$.
		\begin{itemize}
			\item 		If $ \delta_1(u, s) = \bot$, then
			$\delta_1(v, s) = \bot$.
			
			\item If $
			(\delta_1(u, s) \neq \bot ) \wedge 
			(\delta_2(u, s) \neq \bot )	
			\wedge
			(\para(\delta_1(u, s)) \cap 
			\sucall(\delta_2(u, s)) = \emptyset)$ holds, then 
			$
			(\delta_1(u, s) \neq \bot ) \wedge 
			(\delta_2(u, s) \neq \bot )	
			\wedge
			(\para(\delta_1(u, s)) \cap 
			\sucall(\delta_2(u, s)) = \emptyset)$ also holds
		\end{itemize}
	\end{itemize}
	In summary, the second condition 
	of the domination relation between 
	$
	\langle v, \Delta_1(v), \resultR_1(v) \rangle$ and  $\langle v, \Delta_2(v), \resultR_2(v) \rangle$
	also holds.	
\end{proof}

If a tuple $\mathcal{A}$ dominates
another tuple $\mathcal{B}$, 
then by repeatedly applying 
Lemma \ref{l:domination} we know 
it is impossible for
$\mathcal{B}$ to eventually lead to 
a larger WCRT bound than $\mathcal{A}$. Therefore, 
in the graph searching procedure, we can safely discard tuples dominated by others.

Algorithm \ref{al:ewcrt} shows the pseudo-code of the algorithm 
to compute $\max_{\pi\in G} \{ \widetilde{R}(\pi) \}$ by
using
the tuple abstractions to search over $G$ in a \emph{width-first} manner. The algorithm uses
$TS$ to keep the set of tuples at each step, which initially contains a single tuple 
$\langle v_{src}, 
\{ v_{src}\},  c(v_{src}) \}
\rangle$ ($v_{src}$ is the source vertex of $G$). 
Then the algorithm repeatedly generates new tuples for all successors of each tuple in $TS$ by 
(\ref{e:delta}) and (\ref{e:resultR}).
A newly generated tuple is added to $TS$ if there are no other tuples in $TS$ dominating it (line 6 to 7).
{A tuple is removed from $TS$ when all of its successors' tuples have been generated.}
Note that the algorithm always selects a tuple 
that has no predecessors in $TS$ to generate new tuples (line 3), so the searching procedure is width-first.
Finally, $TS$ only contains tuples associated with the sink vertex $v_{snk}$.
By Lemma \ref{l:resultR} we know, 
the $\resultR$ of each tuple 
in the final $TS$ equals the 
$\widetilde{R}(\pi)$ of the corresponding path $\pi$, so 
the maximal $\resultR$ among these final tuple are the desired  $\max_{\pi\in G} \{ \widetilde{R}(\pi) \}$.
By the above discussions, we can conclude the correctness of our algorithm:
\begin{theorem}
	The return value of Algorithm \ref{al:ewcrt} equals $\max_{\pi\in G} \{ \widetilde{R}(\pi) \}$.
\end{theorem}

\begin{algorithm}
	\caption{Pseudo-code for computing $\max_{\pi\in G} \{ \widetilde{R}(\pi) \}$ }\label{al:ewcrt}
	\begin{algorithmic}[1]
		\STATE $TS = \{\langle v_{src}, 
		\{ v_{src}\},  c(v_{src}) 
		\rangle \} $
		
		\WHILE {($  \exists \langle v,  \Delta, \resultR \rangle \in TS : v \neq v_{snk}  $) }
		\STATE Select a tuple $ \langle v, \Delta, \resultR \rangle$ in $TS$  
		s.t. 
		\vspace{-2mm}
		$$\nexists \langle v', \Delta', \resultR' \rangle \in TS : v' \in \pred(v) 
		\vspace{-2mm}
		$$
		\FOR {(each $v' \in \sucd(v)$)}
		\STATE Compute $\langle v'\!, \Delta', \resultR' \rangle$ using $\langle v, \Delta, \resultR \rangle$ by (\ref{e:delta}) and (\ref{e:resultR}) 
		\IF{$(\nexists  \langle {v'}\!, {\Delta^*}\!, {\resultR^*} \rangle \!\!\in\!\! TS \!\!:\langle {v'}\!, \!{\Delta^*}\!, \!{\resultR^*} \rangle \!\succcurlyeq \! \langle v'\!, \! \Delta', \! \resultR' \rangle )$} \label{line:6}
		\STATE $TS = TS \cup \{ \langle v'\!, \Delta'\!, \resultR' \rangle  \}$ 
		\ENDIF \label{line:8}
		\ENDFOR
		
		\STATE  $TS = TS \setminus  \{\langle v, \Delta, \resultR \rangle \} $	 
		
		\ENDWHILE
		\RETURN $\max\{\resultR | \langle v, \Delta, \resultR  \rangle \in TS \}	$
	\end{algorithmic}
\end{algorithm}

The time complexity of Algorithm \ref{al:ewcrt}
depends on the total number of tuples generated in the computation procedure.
As a tuple is put into $TS$ only if no other tuples in $TS$ dominate it, the total number of tuples have ever been recorded in $TS$ in bounded by $O(|V|^{|S|+1})$  (at most $|V|$ different vertcies, $O(|V|^{|S|})$ different values for 
$\Delta$, and only a single $\resultR$ kept for the same $v$ and $\Delta$). Each tuple in $TS$ can generate no more than $|V|$ new tuples, so the total number of tuples ever been generated is bounded by $O(|V|^{|S|+2})$, so the overall time complexity of  Algorithm \ref{al:ewcrt} is $O(|V|^{|S|+2})$.
%
%
%

%
%
%
%

%
%

%
%
%
%
%
%
%
%

\section{Evaluation}\label{s:evaluation}

\begin{figure} 
	\centering 
	\subfigure[Accept. ratio with changing $U$]{\label{fig:accu} 
		\includegraphics[scale=0.152]{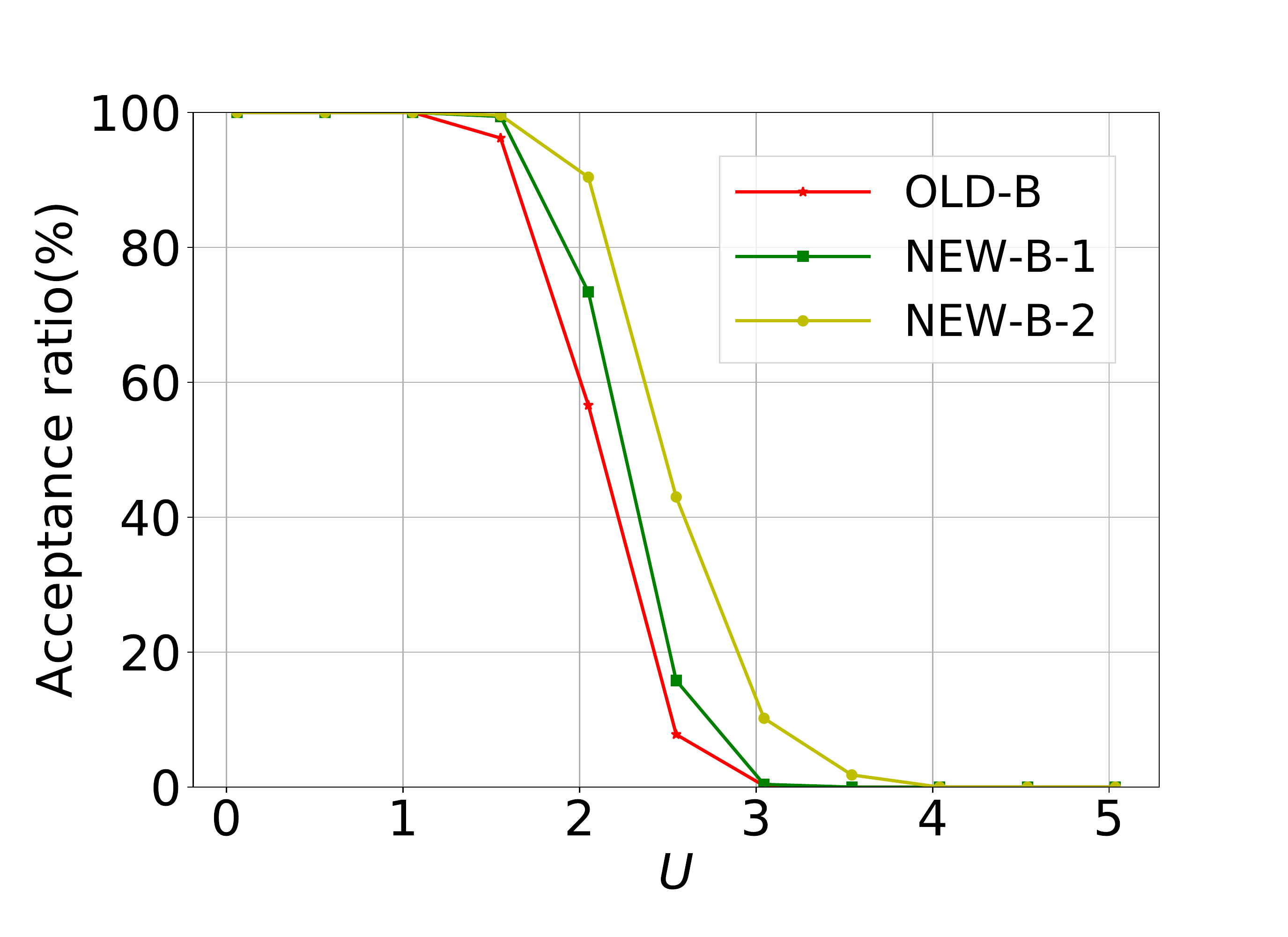}}  
	\subfigure[Norm. bound with changing $U$]{\label{fig:rtru} 
		\includegraphics[scale=0.152]{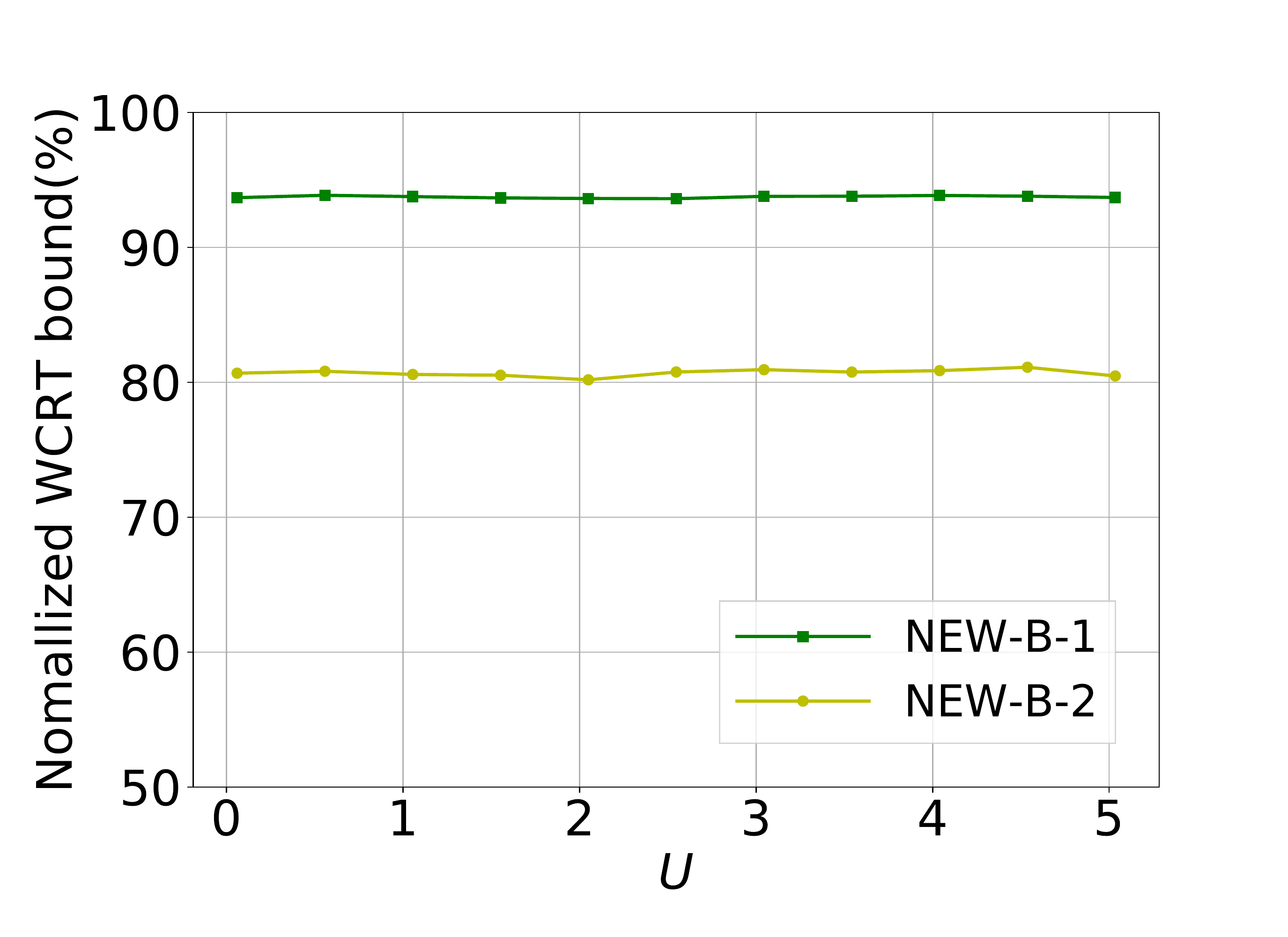}}
	
	\subfigure[Accept. ratio with changing $|V|$]{\label{fig:accN} 
		\includegraphics[scale=0.152]{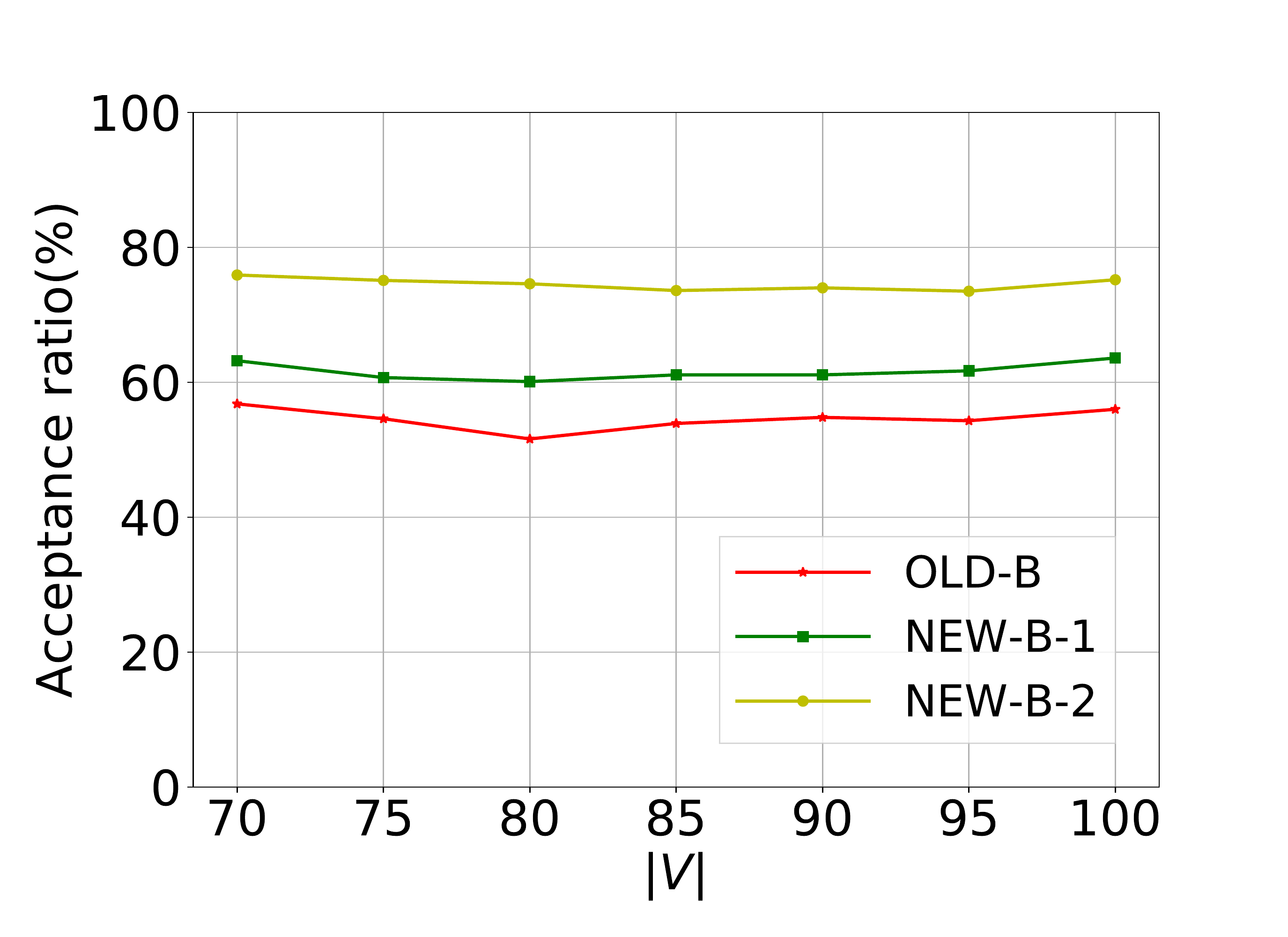}}  
	\subfigure[Norm. bound with changing $|V|$]{ \label{fig:rtrN} 
		\includegraphics[scale=0.152]{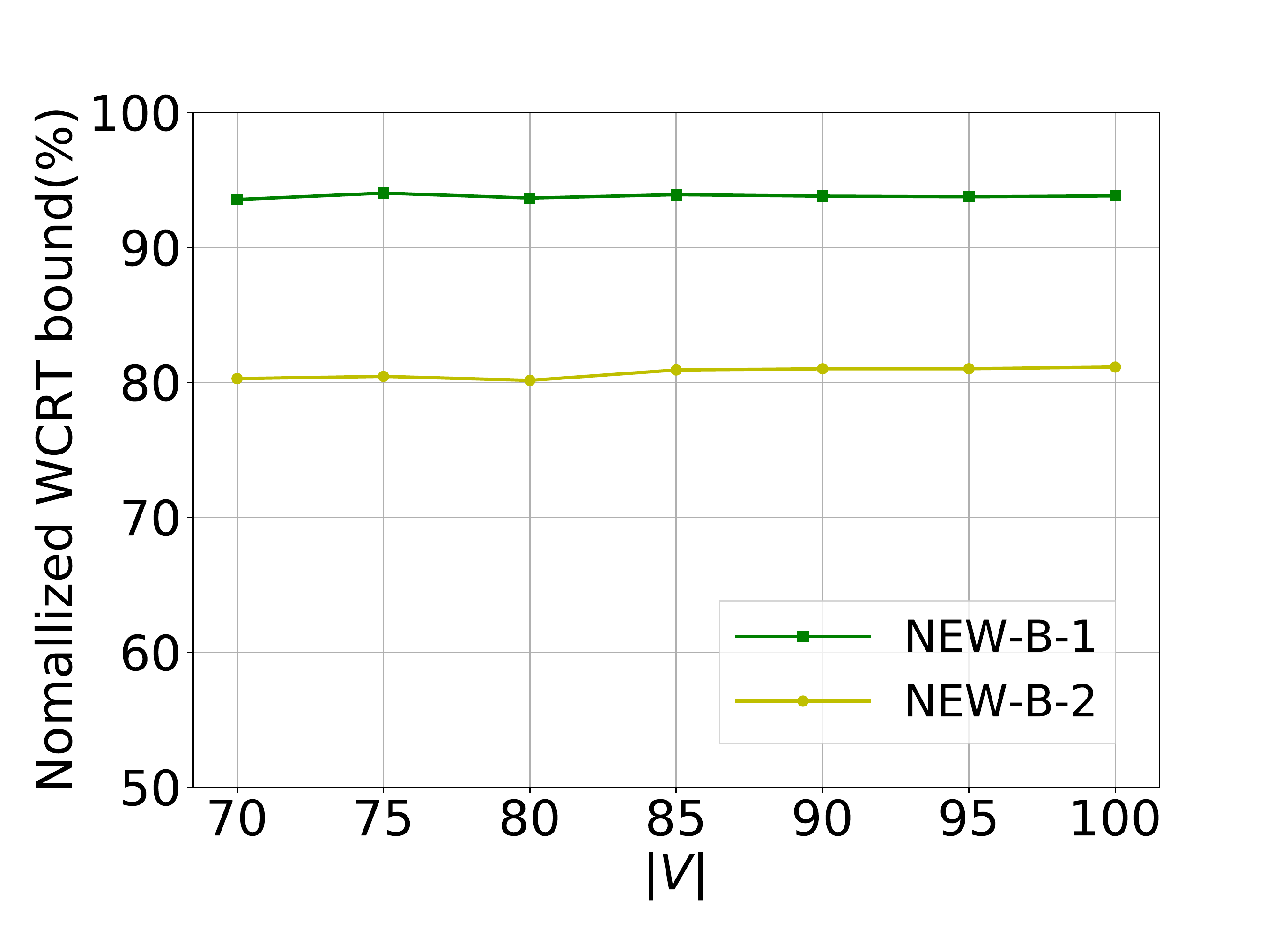}}
	
	\subfigure[Accept. ratio with changing $pr$]{\label{fig:accpr} 
		\includegraphics[scale=0.152]{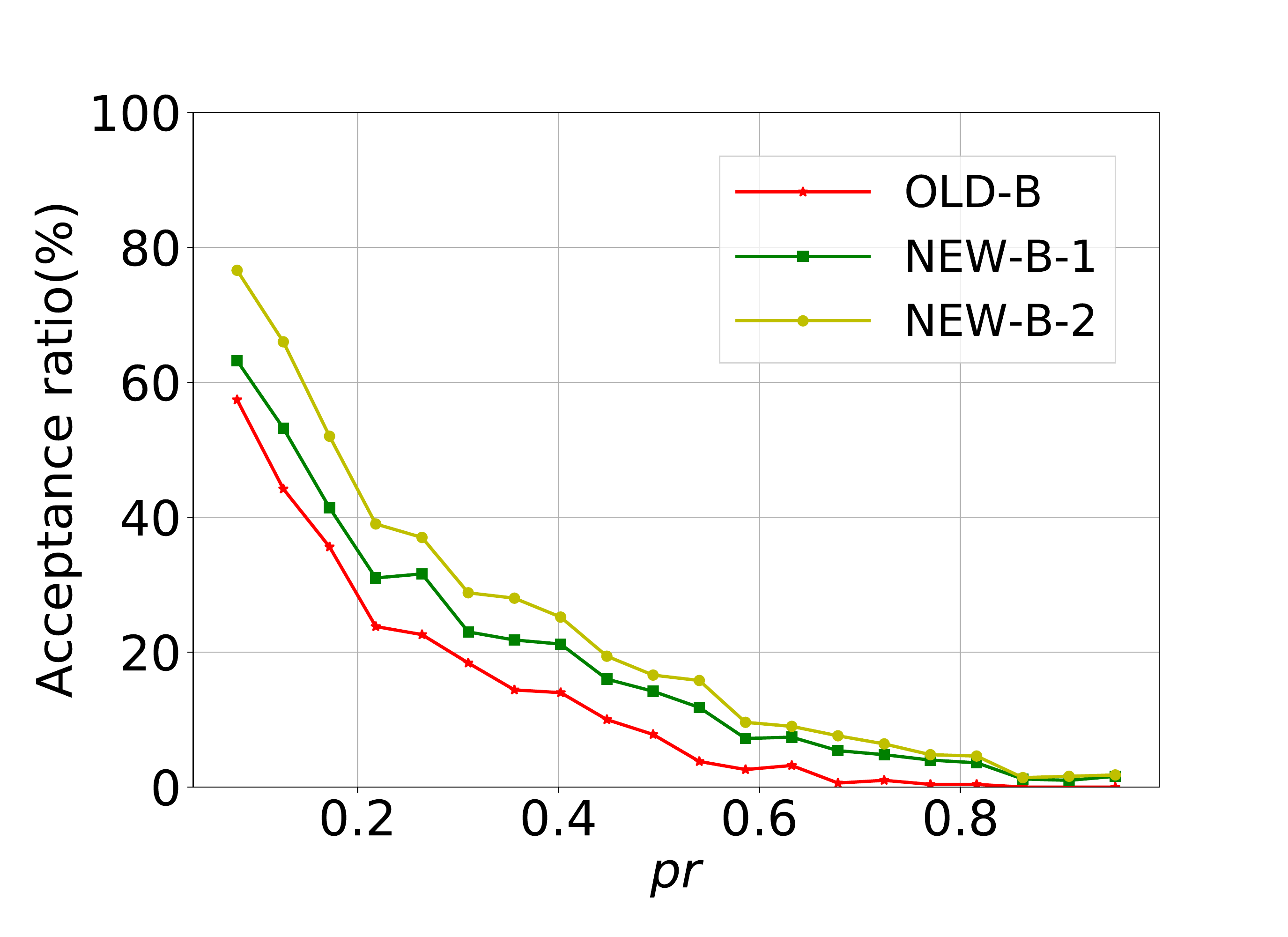}}  
	\subfigure[Norm. bound with changing $pr$]{  \label{fig:rtrpr} 
		\includegraphics[scale=0.152]{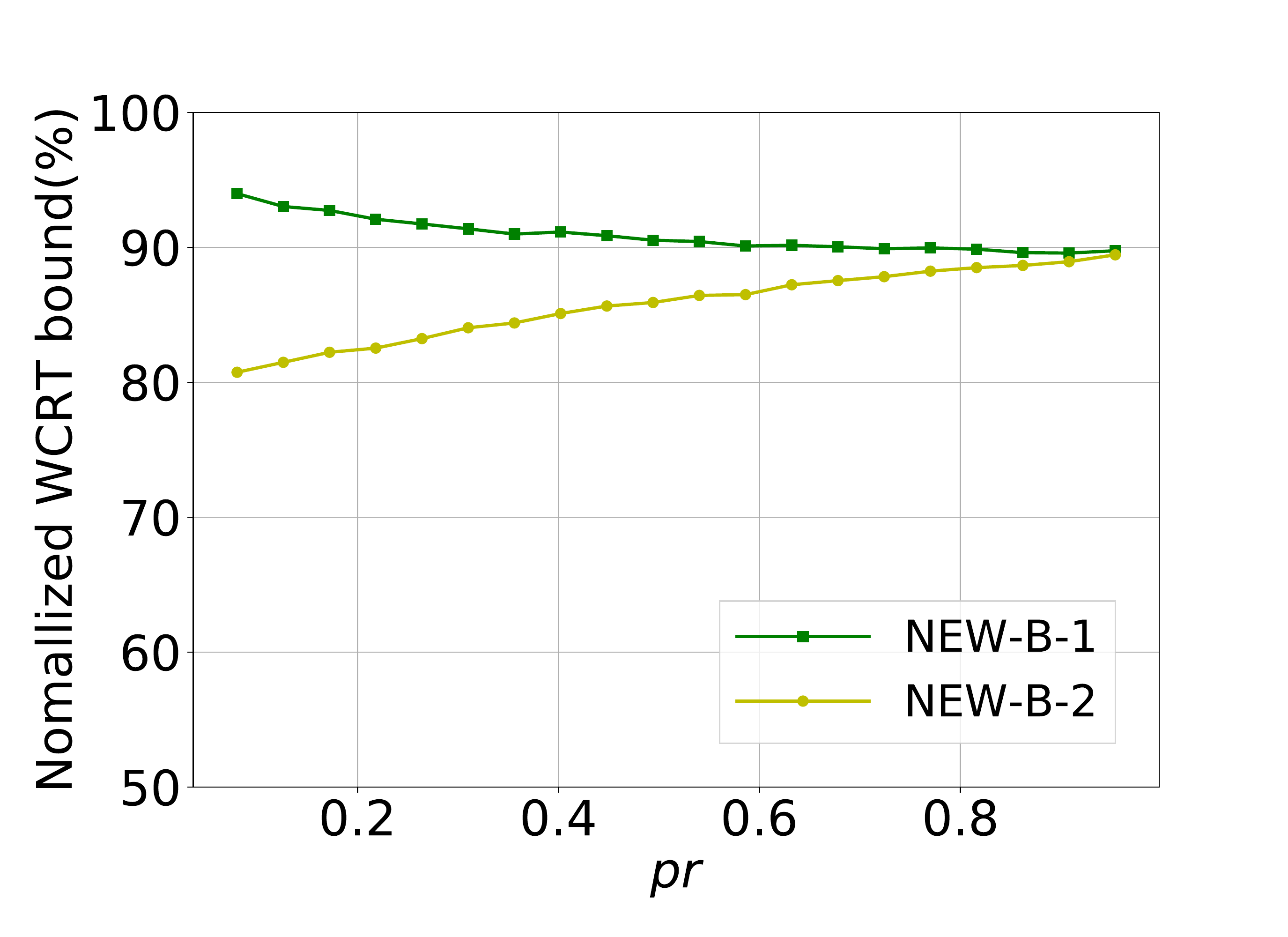}}
	
	\subfigure[Accept. ratio with changing $|S|$]{\label{fig:accpn}  
		\includegraphics[scale=0.152]{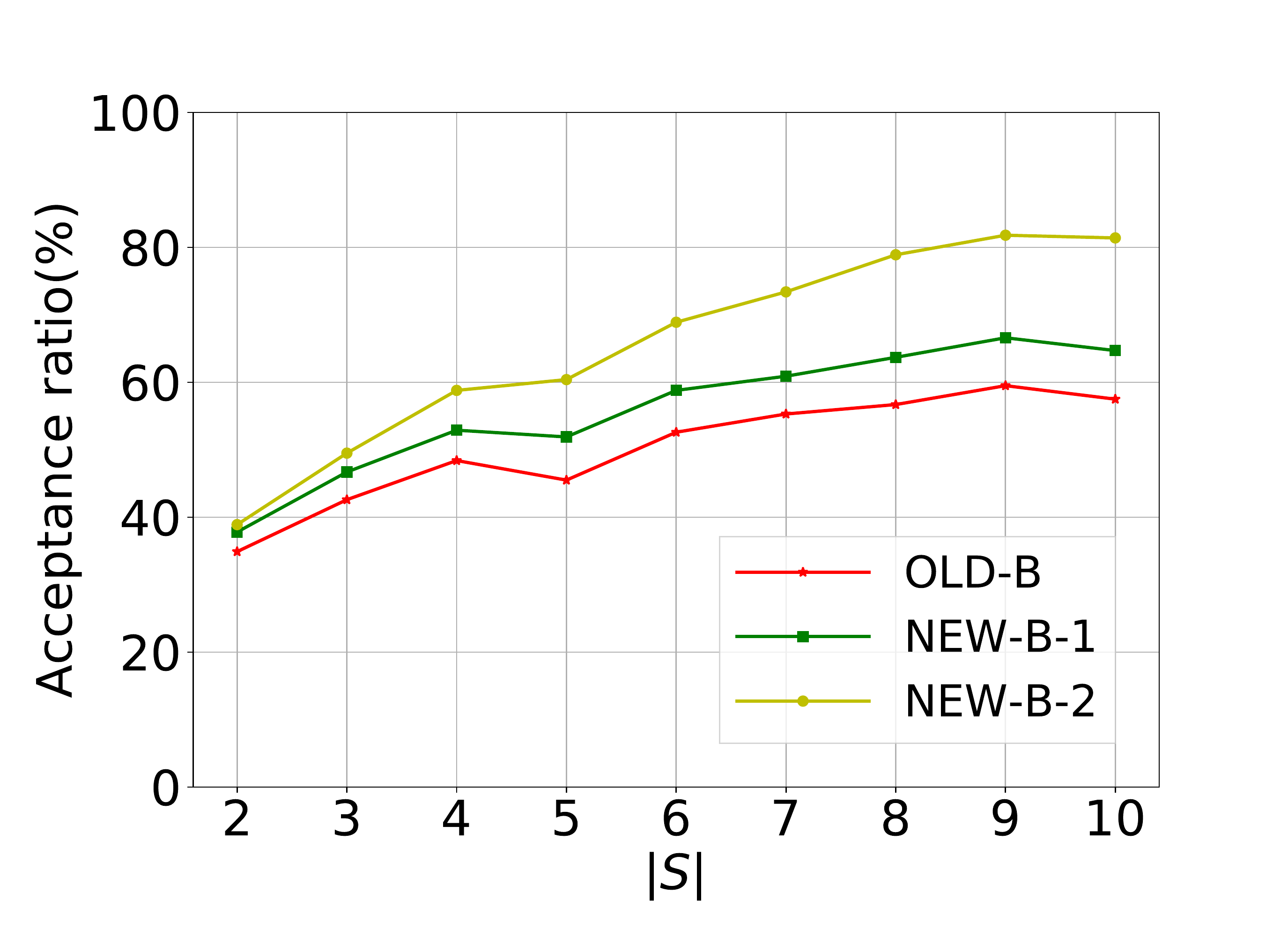}}  
	\subfigure[Norm. bound with changing $|S|$]{ \label{fig:rtrpn}
		\includegraphics[scale=0.152]{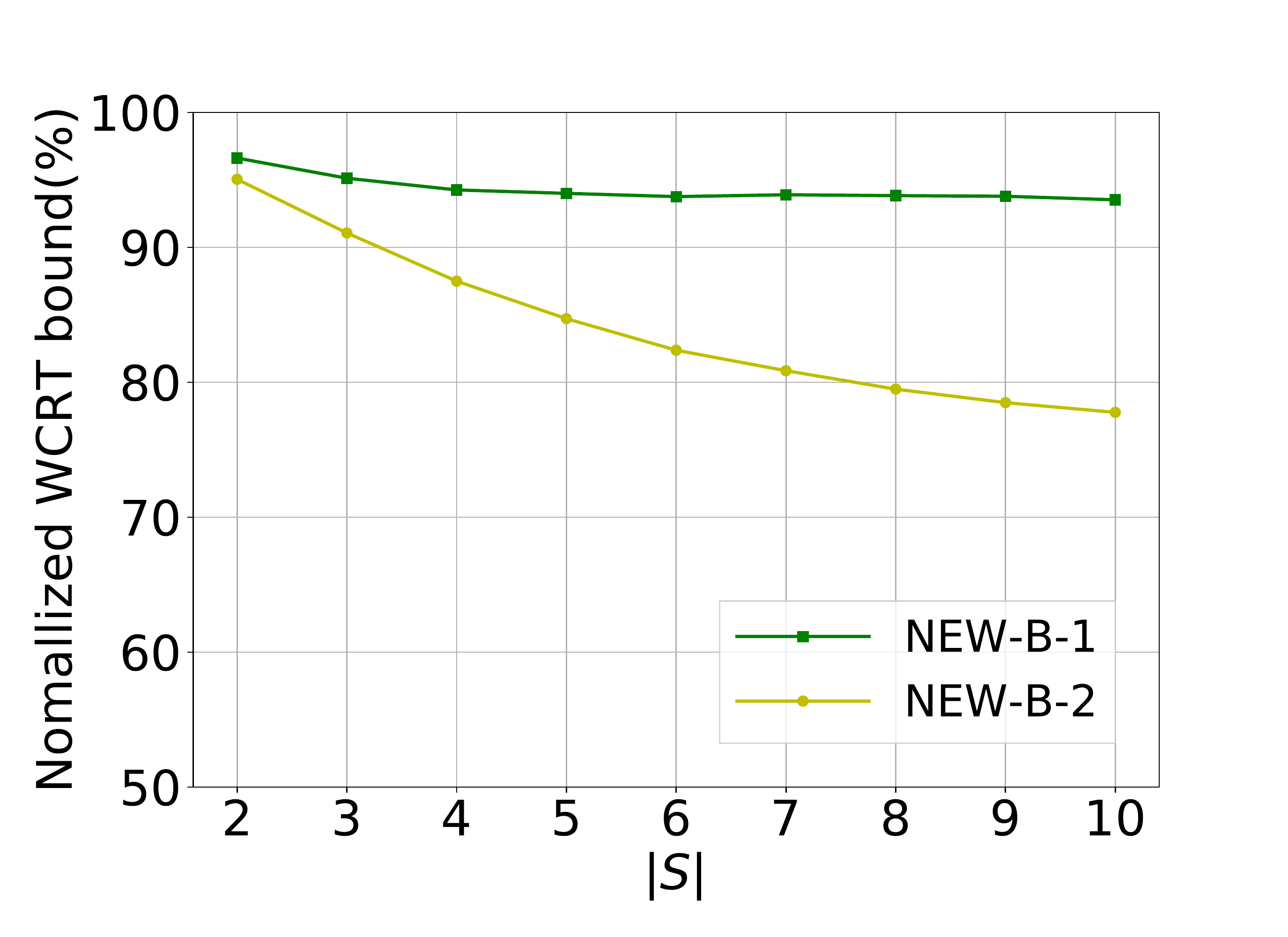}}	
	
	\caption{Comparison of analysis precision of different bounds.} 
	\label{fig:precise} 
\end{figure}

In this section, we experimentally evaluate the performance
of our proposed analysis methods in terms of both precision and efficiency. We compare the following WCRT bounds in the experiments:
\begin{itemize}
	\item \textbf{OLD-B}: the baseline bound in \cite{Jaffe1980} (i.e., Theorem \ref{th:bwcrt}).
	\item \textbf{NEW-B-1}: our first new bound in Theorem \ref{th:newwcrt1}.
	\item \textbf{NEW-B-2}: our second new bound in Theorem \ref{th:wcrt}.
\end{itemize}

We assume the typed DAG tasks are periodic with a period $100$ with implicit deadlines (thus we evaluate not only the WCRT bound but also
the schedulability of each task).
We first define a default parameter setting,
and then tune different parameters to evaluate the performance  of the methods regarding  different parameter changing trends.
The default parameter setting is defined as follows:
\begin{itemize}
	\item The number of types $|S|$ is randomly chosen in
	the range $[5, 10]$, and the number of cores $M_s$ of each type $s$ is randomly chosen in $[2, 11]$.	
	\item The DAG structure of the task is generated by the method proposed in \cite{Cordeiro2010}, where the number of vertices $|V|$ is randomly chosen in the range $[70, 100]$ and 	
	the parallelism factor $p_r$ is randomly chosen in $[0.08,0.1]$ (the larger $p_r$, the more sequential is the graph).
	\item The total utilization $U$ of the typed DAG task is randomly chosen in $[1, 3]$, and thus the total WCET of 
	the task $vol(G) = U \times 100$. 
	\item We use the Unnifast method \cite{Bini2005} to distribute the total WCET to each individual vertex.
	\item Each vertex is randomly assigned a type in $S$.
\end{itemize}

\begin{figure} 
	\centering 
	\subfigure[Analysis time with changing $U$ ]{\label{fig:effu} 
		\includegraphics[scale=0.152]{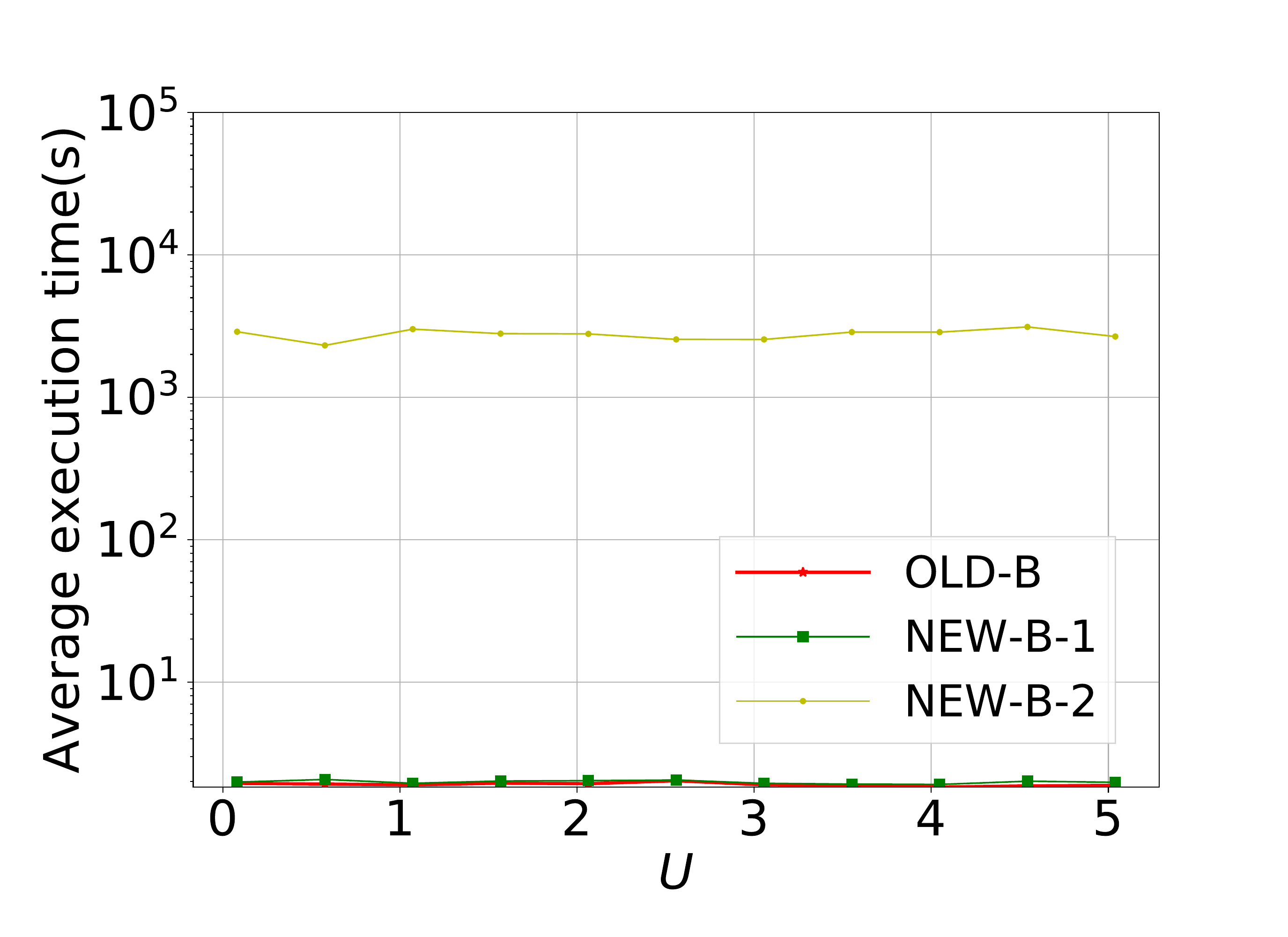}}  
	\subfigure[Analysis time with changing $pr$ ]{\label{fig:effpr} 
		\includegraphics[scale=0.152]{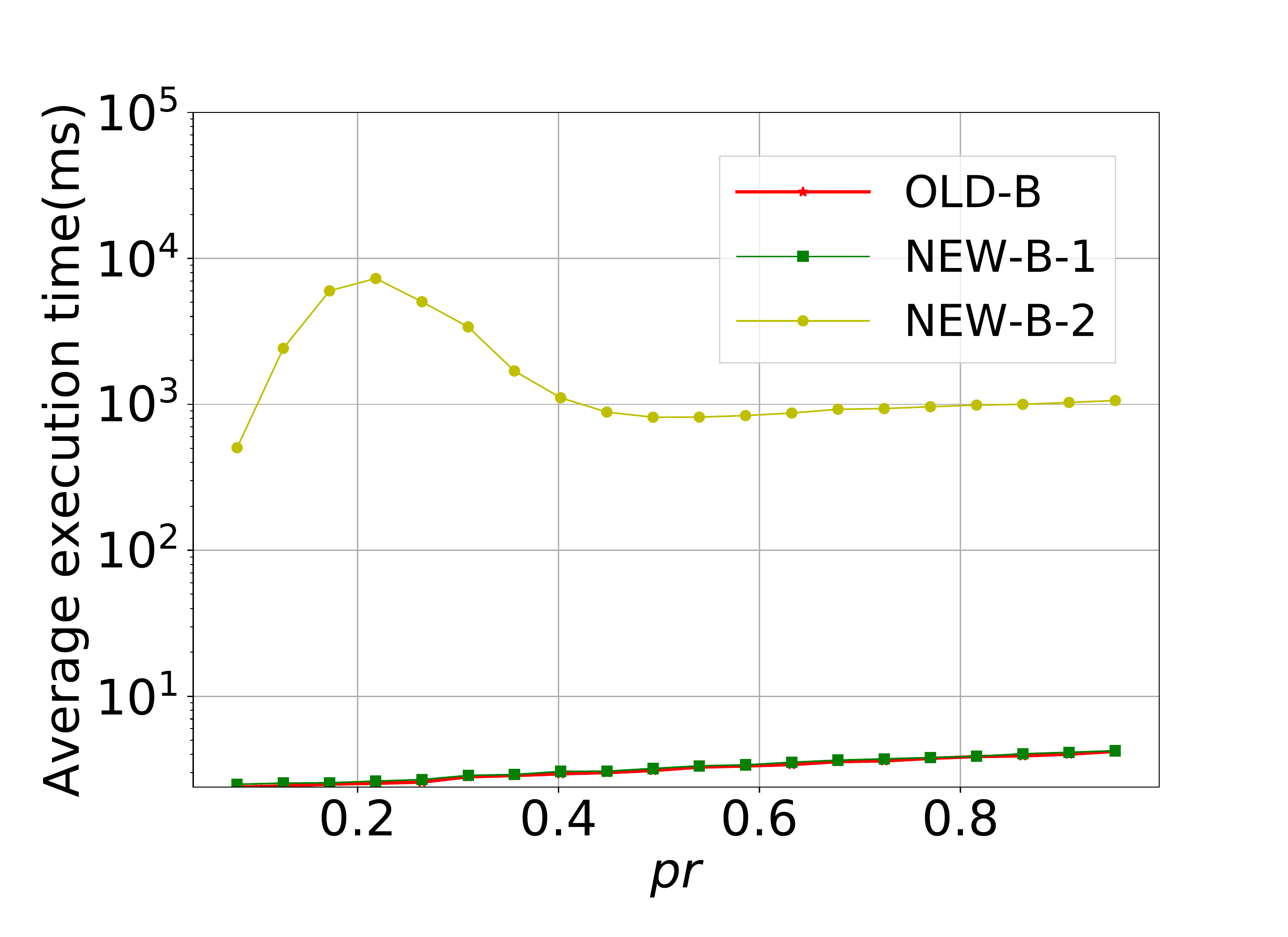}}  
	\subfigure[Analysis time with changing $|S|$]{ \label{fig:effpn} 
		\includegraphics[scale=0.152]{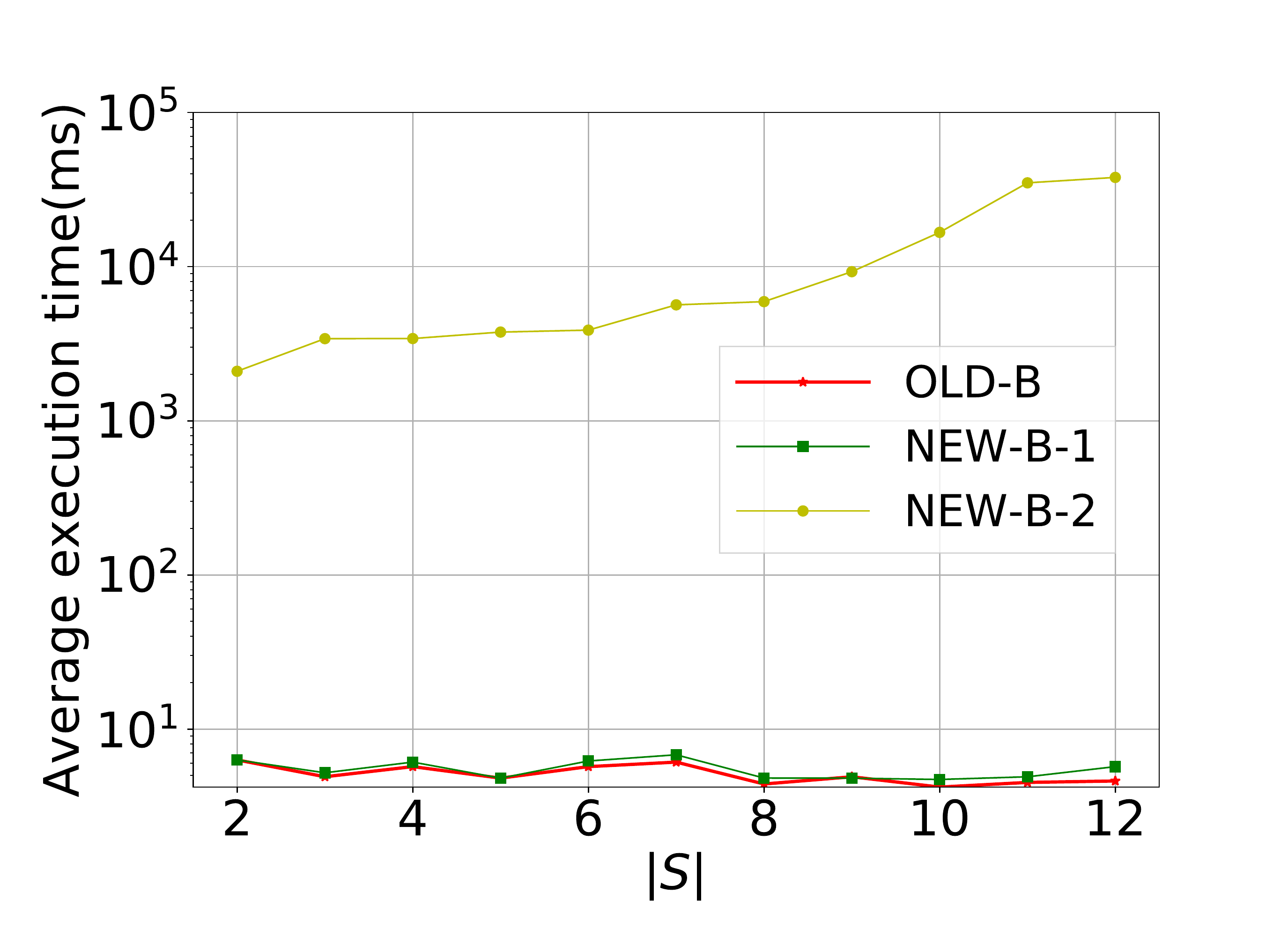}}
	\subfigure[Analysis time with changing $|V|$]{ \label{fig:effN} 
		\includegraphics[scale=0.152]{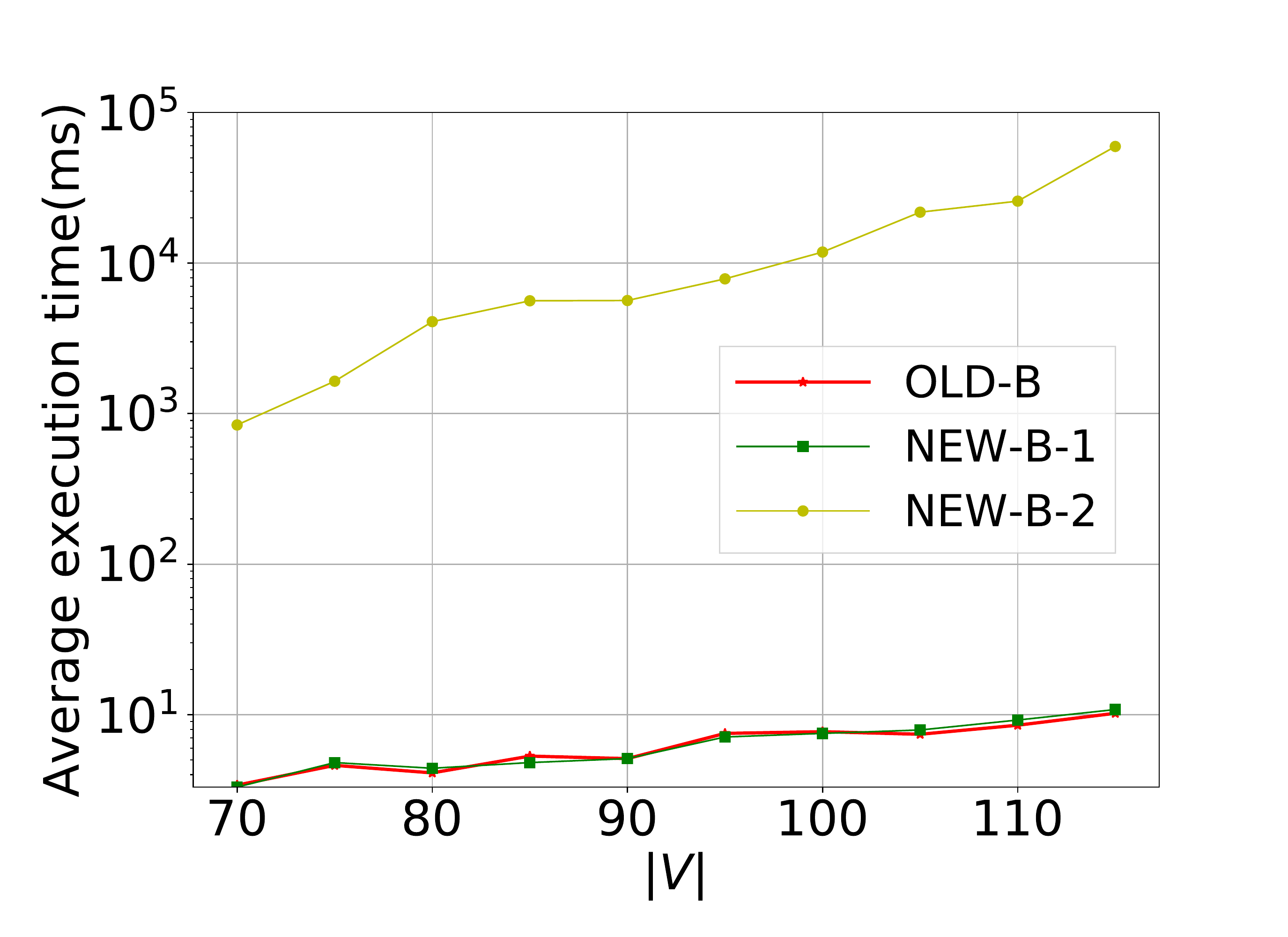}} 
	\caption{Comparison of analysis efficiency of different bounds.} 
	\label{fig:eff} 
\end{figure}

Figure \ref{fig:precise} show the evaluation results for analysis precision with different parameters changed based on the default setting. 
Figure \ref{fig:accu} and \ref{fig:rtru} show experiment results with changing $U$ (other parameters are the same as the default setting). Each point in Figure \ref{fig:accu} represents the \emph{acceptance ratio} of a WCRT bound, which is defined as the ratio between the number of 
tasks decided to be schedulable by a particular WCRT bound and the total number of tasks generated, with the target utilization $U$ on the axis. Figure \ref{fig:rtru} shows the normalized WCRT bound {by} our new methods (the baseline bound is $100\%$). 
Figure \ref{fig:accN} and \ref{fig:rtrN} are results with changing $|V|$, Figure \ref{fig:accpr} and \ref{fig:rtrpr} are results with changing $pr$,
Figure \ref{fig:accpn} and \ref{fig:rtrpn} 
are results with changing $|S|$. From these experiments we can see that our proposed new bounds \textbf{NEW-B-1} and \textbf{NEW-B-2} consistently outperform \textbf{OLD-B} under various parameter settings. 

Figure	\ref{fig:eff} evaluates the computational efficiency of different bounds, with different parameter changed based on the default setting (the same as Figure \ref{fig:precise}). Experiments results show that our first new bound \textbf{NEW-B-1} is almost as efficient as \textbf{OLD-B}, while the computation procedure of \textbf{NEW-B-2} takes much longer time. The lower efficiency of \textbf{NEW-B-2} 
is due to the inherent hardness of its computation problem: exponential with respect to $|S|$
and high-order polynomial with respect to $|V|$ when $|S|$ is large.
The computation time of 
\textbf{NEW-B-2} first increases then decreases 
as $pr$ increases because the number of possible paths
is relatively smaller with very small $pr$ (with which the paths are typically very short) and with large $pr$ (with which the
task graphs are more sequential and thus the number of possible paths is not very large). 
The number of different core types 
in realistic heterogeneous multi-cores is usually a very small number. From Figure \ref{fig:effpn} we can see that the analysis procedure can finish in several seconds when the number of types is smaller than $8$, which should be acceptable for most offline design scenarios.

\begin{figure} 
	\centering 
	\label{fig:paths} 
	\includegraphics[scale=0.22]{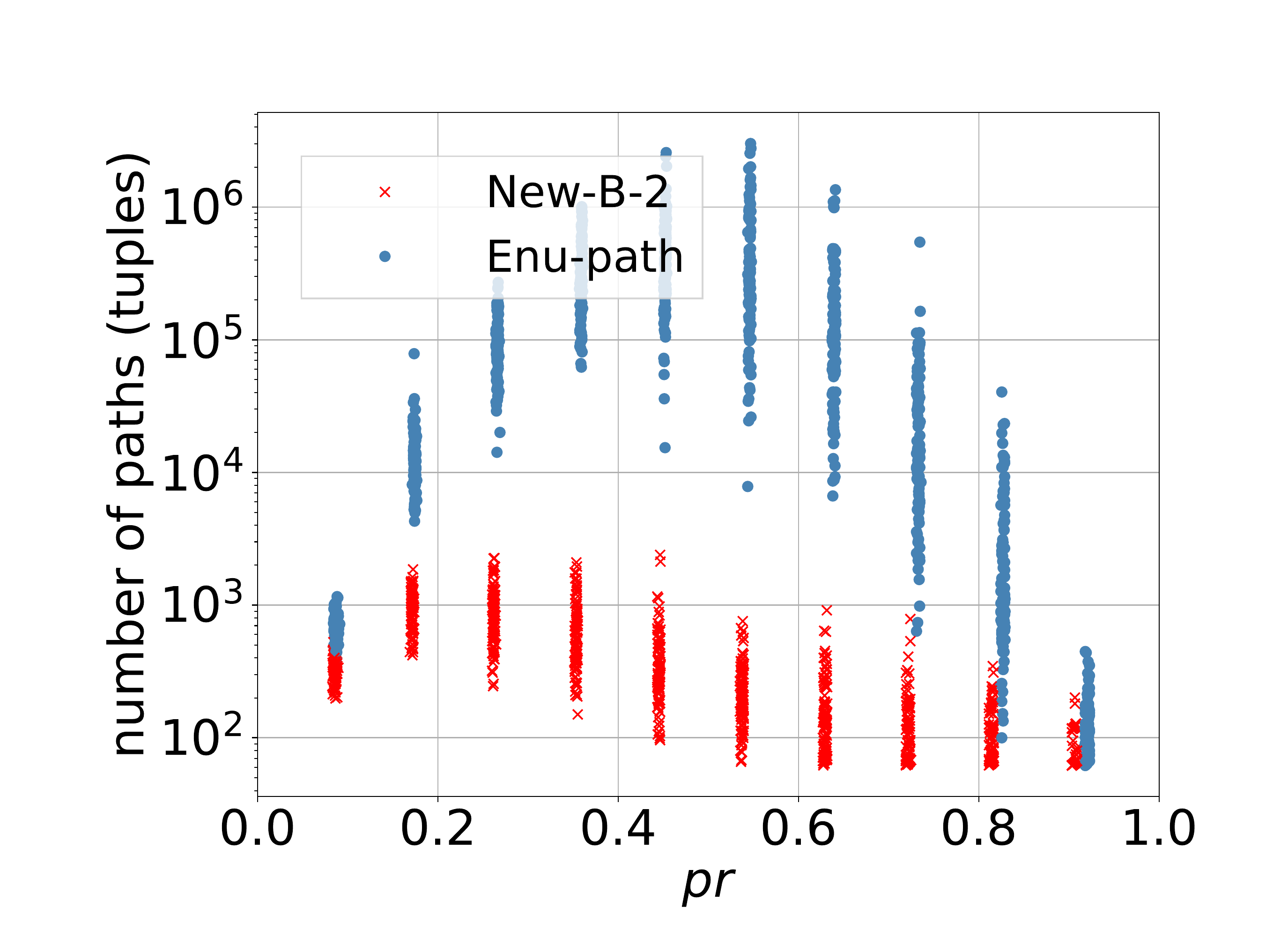}
	\caption{State space reduction by our computation algorithm for \textbf{NEW-B-2}.} 
	\label{fig:path} 
\end{figure}

Figure \ref{fig:path} shows the effectiveness of our proposed algorithm for computing \textbf{NEW-B-2} based on the abstract path representation, by comparing the total number of tuples generated by our algorithm and the total number of paths of the task graph. 
We can see that using our abstraction the searching state space is reduced by typically $3$ to $4$ orders of magnitude. 

\section{Related Work}\label{s:relatedwork}
The most relevant related work is \cite{Jaffe1980},
which developed the baseline WCRT bound \textbf{OLD-B} as stated in 
Section \ref{s:classicmethod}. The analysis techniques of \textbf{OLD-B} are based on the classical work by Graham \cite{Graham1966, Graham1969} for \emph{untyped} DAG (the special case of the model considered in this paper with only one type).
The real-time scheduling problem for multiple \emph{recurring} {untyped} DAG has been intensively studied, with different scheduling paradigm 
including federated scheduling \cite{Bonifaci2013,Baruah2014,Melani2015} and global 
scheduling \cite{Li2014,Baruah2015,Baruah2015a,Baruah2015d,Li2017a}, which are also based on the classical
work by Graham \cite{Graham1966, Graham1969}. Similarly, although we assume a single typed DAG task in this paper, our work would be a necessary step towards the scheduling and analysis of multiple recurring typed DAG tasks.

Yang et al. \cite{Yang2016} studied the scheduling of multiple typed DAG tasks by decomposing each DAG task into a set of independent tasks with artificial release time and deadlines, and then analyzed these decomposed independent tasks by known methods under non-preemptive G-EDF (global earliest deadline first). The work in \cite{Yang2016} is not directly comparable with our work due to the difference in problem models. However, when our method (based on Graham's analysis framework) is extended to the multiple-DAG setting in the future, a comparison between these two families of techniques would be meaningful.

Yang et al. \cite{Yang2016} studied the scheduling of multiple typed
DAG tasks by decomposing each DAG task into a set of
independent tasks with artificial release time and deadlines,
and then analyzed these decomposed independent tasks
by known methods under non-preemptive G-EDF (global
earliest deadline first). The work in \cite{Yang2016} is also applicable to the problem 
studies of this paper which is a special case of the multiple DAG setting in \cite{Yang2016}. 
However, we did not include experimental comparison results between our method and \cite{Yang2016}
as the WCRT bounds yielded by the method in \cite{Yang2016} is much larger than our method.
The transformation to independent tasks in \cite{Yang2016} is to handle the difficulty caused by 
multiple DAGs. Therefore, the comparison between our method and \cite{Yang2016} under the single DAG setting 
will be unfairly in favor to us. We do not want to take of advantage of this to exaggerate the superiority of
our method\footnote{Nevertheless, we provide some comparison results between our method and \cite{Yang2016}
	in  \url{https://github.com/MelindaHan/Typedscheduling} for readers who are interested.} When our method (based on Graham’s
analysis framework) is extended to the multiple-DAG setting
in the future, a comparison between these two families
of techniques would be meaningful.

The scheduling problems with different workload-processor binding restrictions have been studied in the past, including the 
inclusive processing set restriction \cite{Li2015,Li2016,Jia2015,Li2017} {where for any two processor sets assigned to two tasks, one set must be the subset of the other}, the interval processing set restriction \cite{Karhi2013,Karhi2014} {where any two processing sets may have the same machines as the other but one is not a subset of the other} and the tree-hierarchical processing set restrictions \cite{Huo2010,Epstein2011} {where each machine is described as a vertex of a tree and a processing set is a path from root to a leaf}.
These models are all different from our work.
On the other hand, much work has been done 
on scheduling of \emph{independent} tasks on heterogeneous multiprocessor with workload-processor binding restrictions \cite{Huo2010,Lee2011,Xu2015}, and \cite{Leung2008, Leung2016} provided comprehensive surveys for this line of work.

\cite{Jansen1995, Han1998}
studied a special case of the problem model of this paper, which assumes (1) two types of machines, (2) chain-type precedence constraints (3) unit execution time for each scheduling unit.
\cite{Jansen1994} studied the special case where
the task graph structures are either in-trees, out-trees or disjoint unions of chains.
\cite{Li} considered a special case with only two types of machines and assume the execution time of each scheduling unit to be constant so that they can generate a static scheduling list offline (which suffers timing anomalies if applied to the problem model in this paper where each task may execute shorter than its WCET).

\section{Conclusion}\label{s:con}

This paper derives WCRT bounds for typed DAG parallel tasks on 
heterogeneous multi-cores, where the workload of each vertex in the DAG is bound to execute on a particular type of cores.
The only known WCRT bound for this problem is grossly pessimistic and suffers the \emph{non-self-sustainability} problem (a successful design may degrade to be unsuccessful when the parameters become better). We propose two new WCRT bounds to address these problems.
The first new bound is more precise without increasing the time complexity and solves the \emph{non-self-sustainability} problem.
The second new bound explores 
more detailed task graph structure information to greatly improve the precision, but is computationally more expensive. We prove the strong NP-hardness of the computation problem for the second bound, and develop an efficient algorithm which has polynomial time complexity if the number of types is a constant. In next step we will extend the results of this paper to the general setting of multiple recurring typed DAG tasks and compare it with our families of techniques such as the decomposition based approach \cite{Yang2016}. Another possible direction for our future work is to develop more efficient algorithms that compute \textbf{NEW-B-2} approximately, which hopefully can give bounds reasonably close to those computed by the algorithm of this paper in much shorter time.

\ifCLASSOPTIONcaptionsoff
  \newpage
\fi



%
\bibliographystyle{plain}
\bibliography{bibtypedscheduling}
%
%

%
%
\begin{IEEEbiography}[{\includegraphics[width=1in,height=1.25in,clip,keepaspectratio]{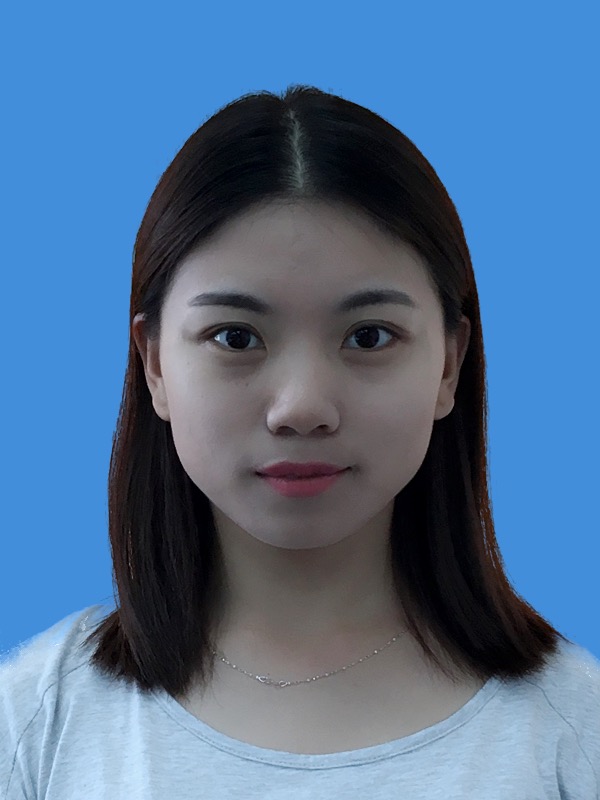}}]{Meiling Han}
as born in Liaocheng, Shandong of China in
1988. She received her master's degree in computer applications technology from Shenyang Agricultural University, China in 2013, she is a Ph.D. candidate at the School of Computer Science and Engineering of Northeastern University, China.
Her research interests are broadly in embedded real-time systems, especially the real-time scheduling on multiprocessor systems.
\end{IEEEbiography}

\begin{IEEEbiography}[{\includegraphics[width=1in,height=1.25in,clip,keepaspectratio]{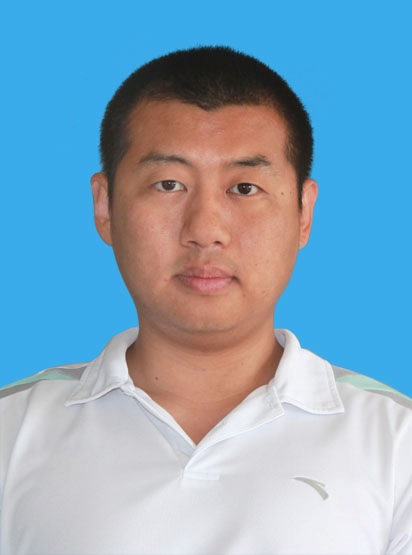}}]{Nan Guan}
	is currently an assistant professor at the Department of Computing, The Hong Kong Polytechnic University. Dr Guan received his BE
	and MS from Northeastern University, China in 2003 and 2006 respectively, and a PhD from Uppsala University, Sweden in 2013. Before joining PolyU in 2015, he worked as a faculty member in Northeastern University, China. His research interests include real-time embedded systems and cyber-physical systems. He received the EDAA Outstanding Dissertation Award in 2014, the Best Paper Award of IEEE Real-time Systems Symposium (RTSS) in 2009, the Best Paper Award of Conference on Design Automation and Test in Europe (DATE) in 2013.
\end{IEEEbiography}


\begin{IEEEbiography}[{\includegraphics[width=1in,height=1.25in,clip,keepaspectratio]{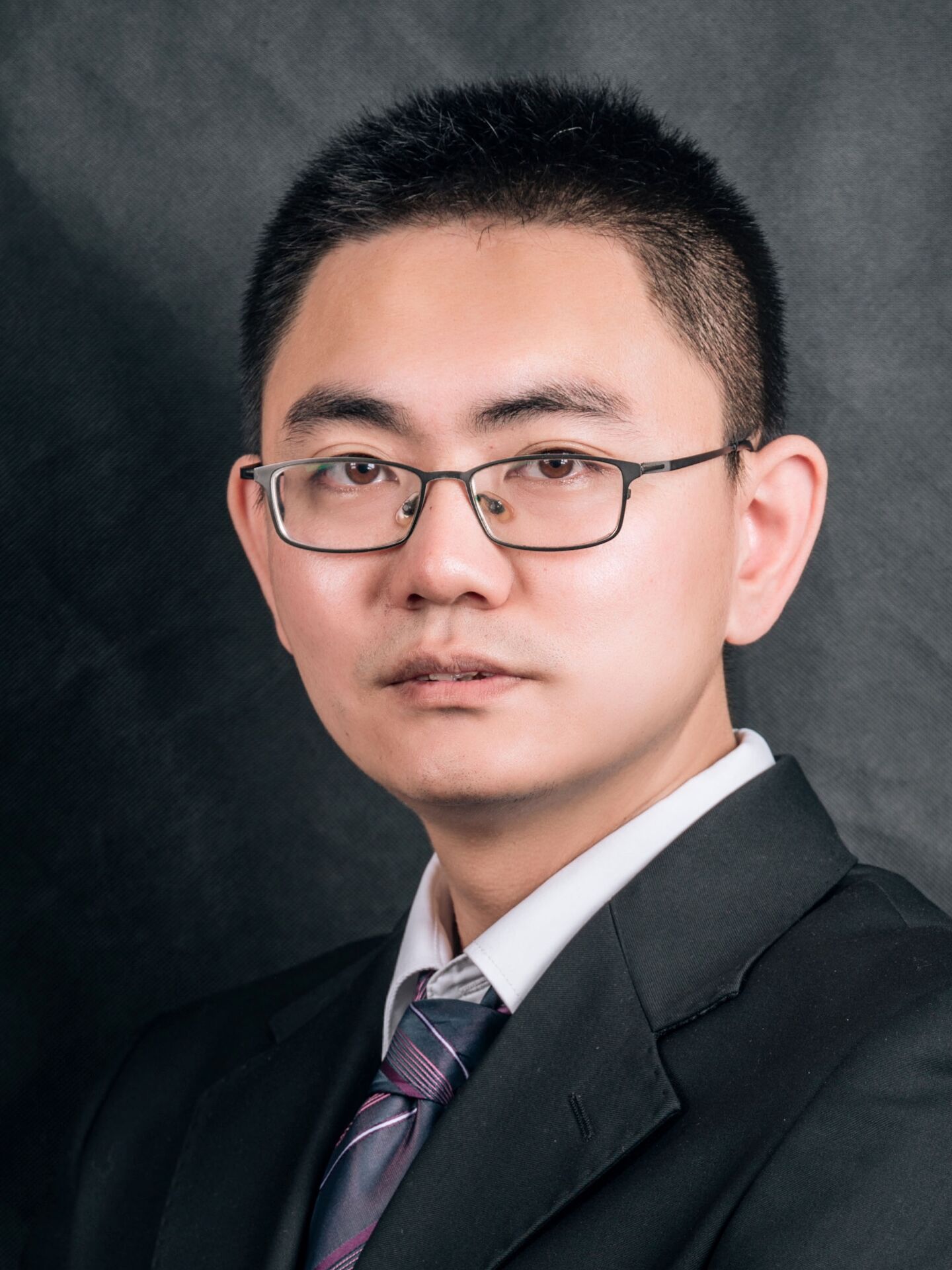}}]{Jinghao Sun}
	received the MS and PhD degree in computer science from Dalian University of Technology in 2012. He is an associated professor at Northeastern University, China. He was a postdoctoral fellow in the Department of Computing at Hong Kong Polytechnic University between 2016 to 2017, working on scheduling algorithms for multi-core real time systems. His research interests include algorithms, schedulability analysis and optimization methods. 
\end{IEEEbiography}

\begin{IEEEbiography}[{\includegraphics[width=1in,height=1.25in,clip,keepaspectratio]{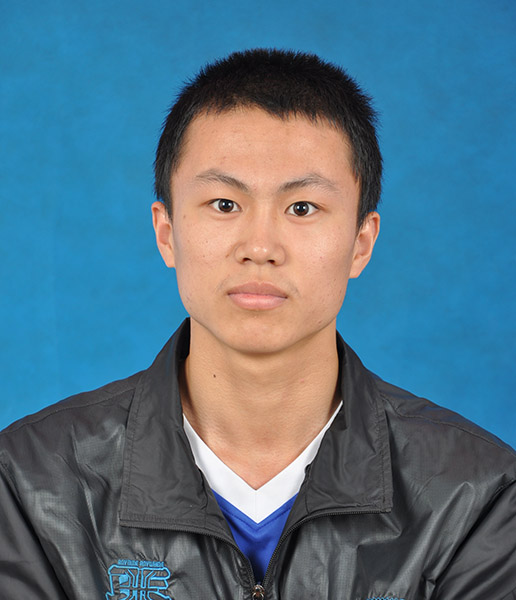}}]{Qingqiang He}
	received the BS degree in computer science and technology from Northeastern University, China, in 2014, and the MS degree in computer software and theory from Northeastern University, China, in 2017. Now he is working as a research assistant in Hong Kong Polytechnic University. His research interests include embedded real-time system, real-time scheduling theory, and distributed leger.
\end{IEEEbiography}

\begin{IEEEbiography}[{\includegraphics[width=1in,height=1.25in,clip,keepaspectratio]{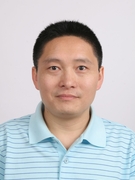}}]{Qingxu Deng}
received his Ph.D. degree in computer science from Northeastern University, China, in 1997. He is a professor of the School of Computer Science and Engineering , Northeastern University, China, where he serves as the Director of institute of Cyber Physical Systems. His main research interests include Cyber-Physical systems, embedded systems, and real-time systems.
\end{IEEEbiography}

\begin{IEEEbiography}[{\includegraphics[width=1in,height=1.25in,clip,keepaspectratio]{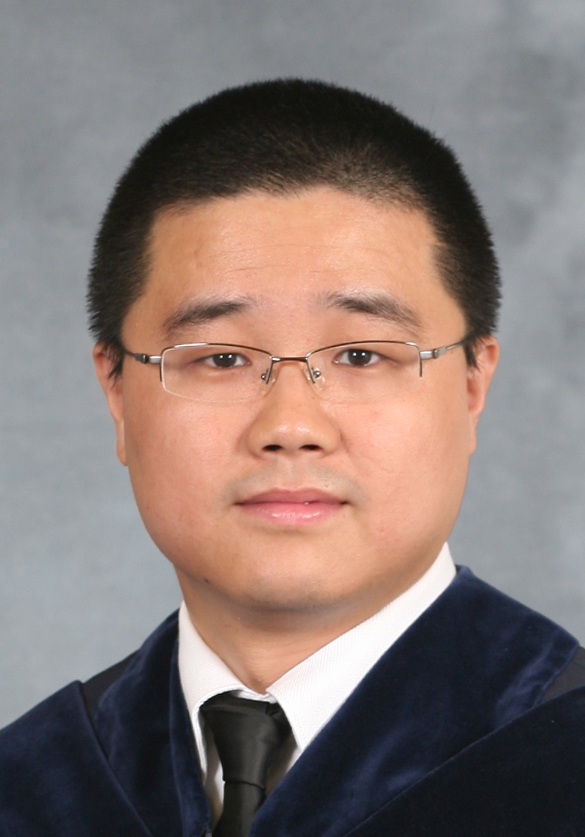}}]{Weichen Liu}
	Weichen Liu (S'07-M'11) is an assistant professor at School of Computer Science and Engineering, Nanyang Technological University, Singapore. He received the PhD degree from the Hong Kong University of Science and Technology, Hong Kong, and the BEng and MEng degrees from Harbin Institute of Technology, China. Dr. Liu has authored and co-authored more than 70 publications in peer-reviewed journals, conferences and books, and received the best paper candidate awards from ASP-DAC 2016, CASES 2015, CODES+ISSS 2009, the best poster awards from RTCSA 2017, AMD-TFE 2010, and the most popular poster award from ASP-DAC 2017. His research interests include embedded and real-time systems, multiprocessor systems and network-on-chip.
\end{IEEEbiography}





\end{document}